\documentclass[prb,reprint,a4paper,citeautoscript]{revtex4-1}

\pdfoutput=1
\usepackage[charter]{mathdesign}
\usepackage{amsmath}

\usepackage[pdftex]{graphicx}
\usepackage{microtype}
\usepackage{bm}

\usepackage[svgnames]{xcolor}
\usepackage[
	colorlinks=True,linkcolor=DarkRed,citecolor=ForestGreen,urlcolor=MediumBlue,
	pdfstartview=FitH,bookmarks=False,pdfpagemode=UseNone
]{hyperref}

\newcommand{\STO}{SrTiO$_3$}
\newcommand{\LAO}{LaAlO$_3$}

\begin{document}

\title{\boldmath Modulation of the superconducting critical temperature due to quantum confinement at the LaAlO$_3$/SrTiO$_3$ interface}

\author{D. Valentinis}
\author{S. Gariglio}
\author{A. F\^{e}te}
\author{J.-M. Triscone} 
\author{C. Berthod}
\author{D. van der Marel} 
\affiliation{Department of Quantum Matter Physics (DQMP), University of Geneva, 24 quai Ernest-Ansermet, 1211 Geneva 4, Switzerland}

\date{September 21, 2017}

\begin{abstract}

Superconductivity develops in bulk doped \STO{} and at the \LAO/\STO{} interface with a dome-shaped density dependence of the critical temperature $T_c$, despite different dimensionalities and geometries. We propose that the $T_c$ dome of \LAO/\STO{} is a shape resonance due to quantum confinement of superconducting bulk \STO{}. We substantiate this interpretation by comparing the exact solutions of a three-dimensional and quasi-two-dimensional two-band BCS gap equation. This comparison highlights the role of heavy bands for $T_c$ in both geometries. For bulk \STO{}, we extract the density dependence of the pairing interaction from the fit to experimental data. We apply quantum confinement in a square potential well of finite depth and calculate $T_c$ in the confined configuration. We compare the calculated $T_c$ to transport experiments and provide an explanation as to why the optimal $T_c$'s are so close to each other in two-dimensional interfaces and the three-dimensional bulk material.
\end{abstract}

\maketitle

\section{Introduction}

The discovery of superconductivity at the interface between \LAO{} and \STO{} (LAO/STO) \cite{Reyren-2007} has spurred fundamental questions regarding the nature of the superconductivity, in particular, the pairing mechanism. A phonon-mediated mechanism \cite{Klimin-2014} has received experimental support \cite{Boschker-2015}, but unconventional scenarios have also been put forward \cite{Stephanos-2011, Scheurer-2015a}. STO has a domelike dependence of $T_c$ on the electron density both in the three-dimensional (3D) bulk material \cite{Schooley-1964, *Schooley-1965, *Koonce-1967, Lin-2014} and in the interfaces \cite{Caviglia-2008, Joshua-2012, Bell-2009, Bert-2012}, with approximately the same value of the maximum $T_c$. Carriers are confined within a thickness of typically 10~nm at low temperature \cite{Reyren-2007, Caviglia-2008, Copie-2009, Dubroka-2010, Gariglio-2015b}, which causes quantum confinement to an effective two-dimensional electron liquid (2DEL). Hartree shifts in the $3d$ shell affect the band structure \cite{Maniv-2015}, and Rashba coupling at the interface may mix singlet and triplet pairing channels \cite{Gorkov-2001, Scheurer-2015a}. The superfluid density in the 2DEL is much lower than the normal-state Hall density, \cite{Bert-2012, Kirtley-2012} suggesting that 2D fluctuations influence the superconducting properties \cite{Caviglia-2008, Benfatto-2009}. Some experiments have reported evidence of magnetic order coexisting with, or phase segregated from, the superconducting phase \cite{Li-2011, Bert-2011, Dikin-2011, Daptary-2017}. Disorder at the interface may also induce phase separation and percolation \cite{Caprara-2012, *Caprara-2013}.

In a broad perspective, the LAO/STO 2DEL may be regarded as an instance of a quasi-2D superconductor produced by the confinement of a 3D superconductor in a film geometry \cite{Bianconi-2014}. Quantum confinement is a well-known strategy for changing the $T_c$, which has been continuously explored theoretically and experimentally since the sixties \cite{Thompson-1963, [][{,} and references therein.]Valentinis-2016b}. The vertical localization of electronic states in the film changes the pairing strength, and the formation of subbands leads to oscillations of $T_c$---often termed shape resonances---as a function of the confinement length and/or carrier density. In this context, the different $T_c$ behaviors in STO and LAO/STO do not rule out a common mechanism. 

Here we propose an interpretation of the dome in LAO/STO as the result of a shape resonance triggered by the confinement of carriers at the STO surface. We support this scenario with a microscopic calculation reproducing both the STO and the LAO/STO superconducting domes with the same pairing model, taking into account measurements of the 2DEL thickness and confinement effects for the latter. According to the new interpretation, the differences between the two domes confirm the commonality of the pairing mechanism for the 2D and 3D cases. LAO/STO offers new opportunities to study shape resonances, thanks to a continuously tunable carrier density, in contrast to metallic thin films, whose thickness can only be changed in increments of 1 unit cell. Our modeling, furthermore, clarifies the roles of the STO light and heavy bands in LAO/STO superconductivity, showing that $T_c$ is controlled almost exclusively by the heavy one.

In order to describe the superconductivity of bulk STO, we first build in Sec.~\ref{model-3D} a minimal microscopic model able to capture the density dependence of $T_c$ with a small number of parameters. In Sec.~\ref{exp}, we discuss recent experimental data by some of us on the LAO/STO 2DEL thickness \cite{Gariglio-2016}. With these data at hand, we can deduce from the measured Hall 2D density an equivalent 3D density for the 2DEL. This allows us to confine the bulk model along one dimension, setting the carrier density to the equivalent 3D density of the 2DEL, and thus to obtain in Sec.~\ref{model-q2D} a prediction for $T_c$ in the confined geometry, which we compare with the measurements at the LAO/STO interface. Our main results are interpreted and discussed in Sec.~\ref{res}; conclusions and perspectives are given in Sec.~\ref{concl}.

\section{Two-band model for bulk STO}\label{model-3D}

\begin{figure}[tb]
\includegraphics[width=\columnwidth]{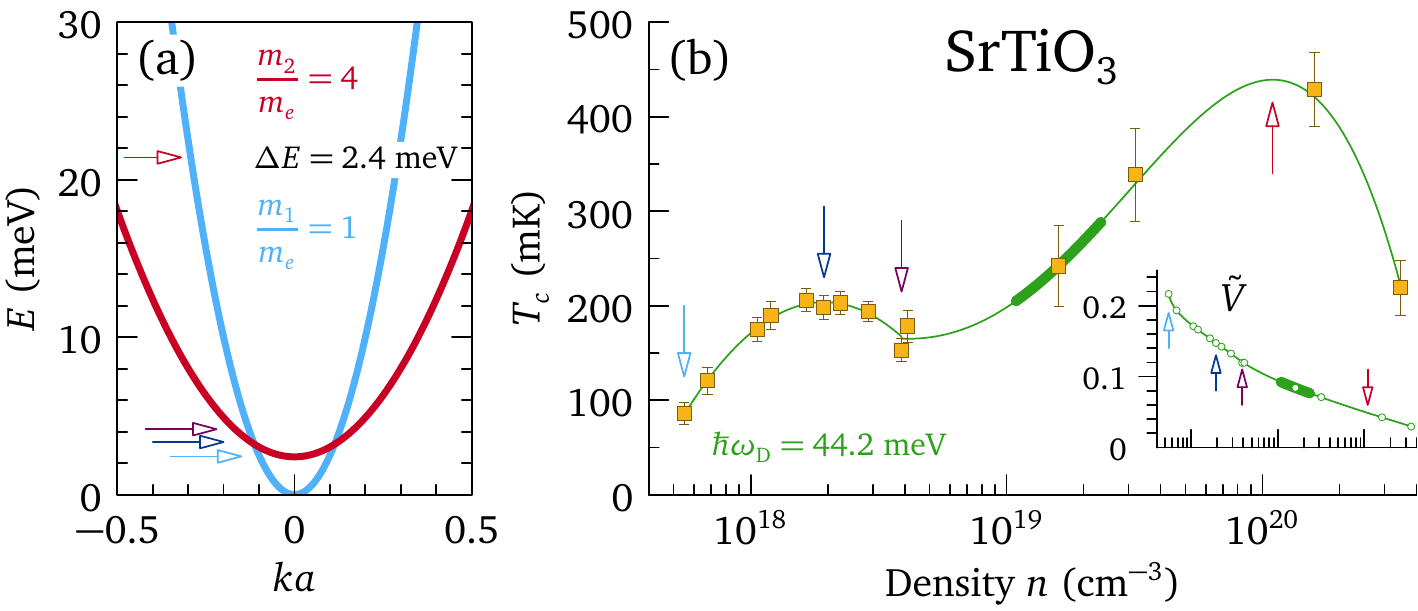}
\caption{\label{fig:fig1}
Weak-coupling model for the density-dependent superconductivity of \STO. (a) Parabolic approximation for the bottom of the conduction band; $\Delta E$ is the band splitting at $k=0$. (b) Data from Ref.~\onlinecite{Lin-2014} (squares with error bars) and calculated $T_c(n)$ curve (solid line). Thick lines in show the range of densities visited at the LAO/STO interface. Inset: Dimensionless interaction strength \cite{Valentinis-2016a} $\tilde{V}=2V[m_e/(2\pi\hbar^2)]^{3/2}\sqrt{\hbar\omega_{\mathrm{D}}}$ reproducing the experimental $T_c$ (circles) and interpolation (solid line). Colored arrows indicate remarkable values of $T_c$ in (b) and the corresponding values of the chemical potential in (a).
}
\end{figure}

Focusing on the essential ingredients, we opt for two parabolic bands and a BCS pairing interaction to describe STO. The low-energy sector of the STO conduction band involves three bands of mostly Ti $3d$ character, split by a crystal field and spin-orbit interaction \cite{vanderMarel-2011, Allen-2013}. We discard one band lying 30~meV above the other two and not occupied at the densities considered in this study. The bare masses of the remaining two bands and their splitting at the $\Gamma$ point are chosen such that the 2D density of states evaluated at $k_z=0$ with the tight-binding and parabolic dispersions agree best. We then renormalize both band masses by a factor of 2, representing the effect of electron-phonon coupling and the emergence of large polarons \cite{Eagles-1969c, vanMechelen-2008, *Devreese-2010}. The resulting band structure is shown in Fig.~\ref{fig:fig1}(a), where the parameters are indicated as well. The pairing mechanism in STO is still debated \cite{Klimin-2014, Edge-2015}. Standard phonon-mediated pairing is questioned because the low density of STO puts it in the antiadiabatic regime $E_{\mathrm{F}}<\hbar\omega_{\mathrm{D}}$, where $E_{\mathrm{F}}$ and $\hbar\omega_{\mathrm{D}}$ are the Fermi and Debye energies, respectively. Leaving alone the origin of pairing, it is believed that STO is amenable to a BCS description with low coupling constants, of the order of 0.1--0.3 \cite{vanderMarel-2011, Allen-2013}. We adopt a local BCS interaction of strength $V$ with a dynamical cutoff $\hbar\omega_{\mathrm{D}}=44$~meV, consistent with the Debye temperature of STO \cite{Ahrens-2007}. The interaction is the same in both bands and we neglect interband coupling, for simplicity. In this model, for $V$ independent of the density $n$, the critical temperature can only increase with increasing $n$ \cite{Fernandes-2013, Valentinis-2016a}, in contradiction with the observations \cite{Schooley-1964, *Schooley-1965, Koonce-1967, Lin-2014}. To explain the dome, $V$ must drop with increasing $n$ \cite{Lin-2014}. We fix the $n$-dependent interaction such that the model reproduces the $T_c(n)$ data from Ref.~\onlinecite{Lin-2014}. This data set is preferred because it covers a broad range of densities, but this choice does not affect any of our conclusions. The resulting interaction decreases monotonically with increasing $n$ as shown in the inset in Fig.~\ref{fig:fig1}(b). We interpolate this dependency to get a continuous parametrization $V(n)$. Figure~\ref{fig:fig1}(b) displays the data from Ref.~\onlinecite{Lin-2014} on top of the continuous $T_c(n)$ curve resulting from that interpolation. The calculation takes full account of the energy-dependent density of states, including the fact that $E_{\mathrm{F}}<\hbar\omega_{\mathrm{D}}$, and uses the self-consistent chemical potential calculated at $T_c$ \cite{Valentinis-2016a}. With this parametrization the heavy band starts to be populated before the first maximum of the double dome, as shown by quantum oscillations \cite{Lin-2014}. A glimpse at the band structure and the value of $\hbar\omega_{\mathrm{D}}$ reveals that the heavy band contributes to the pairing instability even when it is not populated (see arrows in Fig.~\ref{fig:fig1}).\footnote{In fact, the contribution of the heavy band is dominant: if the interaction is switched off in the light band, the calculated $T_c$ changes by less than a percent, while if the interaction in the heavy band is reduced by 20\%, $T_c$ does not exceed 80~mK at any density.}

\section{Experimental determination of the 2DEL thickness in LAO/STO}\label{exp}

Our hypothesis is that the pairing strength at the LAO/STO interface follows the bulk interaction $V(n)$. The carrier density varies across the 2DEL, following the profile of the confinement potential, such that the pairing strength would be, strictly speaking, a function of the position. As the coherence length is large (50--100~nm) compared with the typical confinement length ($\approx 10$~nm), which itself is only marginally larger than the Fermi wavelength \cite{Copie-2009}, we use an average $V$ associated with the average carrier density $n=n_{\mathrm{2D}}/L$. The density integrated along the confinement direction, $n_{\mathrm{2D}}=\int_{-\infty}^{\infty}dz\,n(z)$, is deduced from the Hall sheet conductance as in Ref.~\onlinecite{Caviglia-2008}. For an estimation of the effective 2DEL thickness $L$, we resort to the analysis of the superconducting transitions measured in magnetic fields perpendicular and parallel to the interface \cite{Reyren-2009}. The samples were prepared by pulsed laser deposition \cite{Cancellieri-2010}. For the transport measurements in a dilution cryostat, field-effect devices were realized using the STO substrate as the gate dielectric and Hall bars were photolithographically defined. For each magnetic field, $T_c$ was defined as the midpoint of the transition: in this way the critical fields display a linear temperature behavior in the perpendicular orientation, while in the parallel orientation there is a square root temperature dependence. This confirms the 2D nature of the superconducting state across the whole phase diagram. The angular dependence of the critical fields allows us to determine the in-plane coherence length $\xi(T)$ and the effective thickness $L$ of the superconducting 2DEL \cite{Gariglio-2015a}. From the perpendicular critical field $H_{c\perp}(T)$, we extract $\xi^2(T)=\phi_0/[2\pi\mu_0 H_{c\perp}(T)]$, where $\phi_0$ is the flux quantum. Extrapolating to zero temperature we obtain $\xi(0)$, and using the experimental parallel critical field $H_{c\parallel}(T)$  we arrive at the superconducting thickness $L=\sqrt{3}\phi_0/[\pi\mu_0\xi(0) H_{c\parallel}(0)]$. The in-plane coherence length remains larger than the superconducting thickness for all dopings \cite{Gariglio-2015a, Gariglio-2015b}, which is again consistent with the 2D character of the superconducting 2DEL.

\begin{figure}[tb]
\includegraphics[width=\columnwidth]{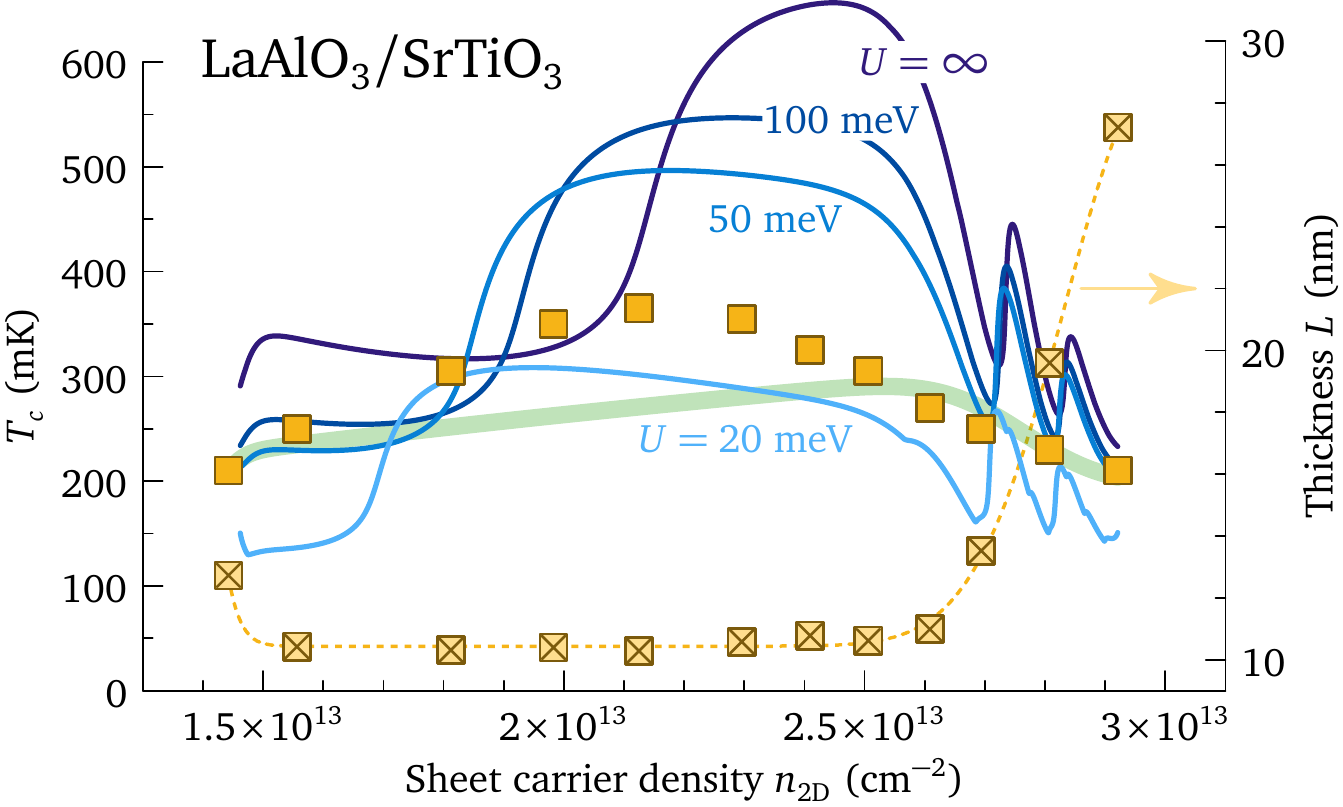}
\caption{\label{fig:fig2}
2DEL critical temperature (squares) and thickness (crossed squares, right scale) measured at the LAO/STO interface as a function of the sheet carrier density. Solid lines show the calculated $T_c$ obtained by confining the model show in Fig.~\ref{fig:fig1} in a square potential well of width $L$ and various depths $U$ as indicated. A continuous interpolation of the thickness as a function of the density (dotted yellow line) was used. For each couple $(n_{\mathrm{2D}},L)$, the pairing interaction is read from the inset in Fig.~\ref{fig:fig1}(b) at the density $n=n_{\mathrm{2D}}/L$. The thick line shows the 3D $T_c$ at density $n$.
}
\end{figure}

Figure~\ref{fig:fig2} displays the thicknesses and critical temperatures measured as a function of the Hall density $n_{\mathrm{2D}}$ \cite{Gariglio-2016}. The dependence $T_c(n_{\mathrm{2D}})$ draws a dome with a maximum at 365~mK. The measured thickness is close to 10~nm except at densities above $2.5\times10^{13}$~cm$^{-2}$, where it increases steadily. Figure~\ref{fig:fig1}(b) shows that the field-effect doping at the interface explores a relatively narrow range of 3D densities (thick line) compared with the chemical doping. The sharp increase in $L$ from $\sim10$ to $\sim30$~nm between 2.5 and $3\times10^{13}$~cm$^{-2}$ is somewhat surprising, especially when contrasted with the density independence of $L$ at lower densities. The variation of $L$ stems from a reduction in $H_{c\parallel}$ by a factor of 3, while $H_{c\perp}$ remains approximately constant. This suggests a qualitative change in the shape of the confining potential at high densities. However, extrinsic effects such as disorder that are unrelated to the 2DEL thickness could also contribute to the suppression of the parallel critical field. It should be noted that the effective thickness comes from a Ginzburg-Landau model with uniform order parameter in a deep square well \cite{Tinkham-1996}. The solution of a more realistic description using a self-consistent potential well cannot be obtained in analytically closed form and is not pursued here for this reason.

\section{Critical temperature of STO carriers confined in a square well}\label{model-q2D}

Knowing the equivalent 3D density $n=n_{\mathrm{2D}}/L$ for each measured $n_{\mathrm{2D}}$, we can determine the pairing strength $V(n)$ in the 2DEL. We model the confinement by taking the band structure of Fig.~\ref{fig:fig1}(a) into a square potential well of width $L$ and depth $U$, following Ref.~\onlinecite{Valentinis-2016b}. This leads to the formation of 2D subbands at energies that depend on $L$ and $U$. All subbands are coupled by the pairing interaction, which, furthermore, gets renormalized by the confinement and becomes a function of the band and subband indices via the bound wave functions. The potential well at the LAO/STO interface is certainly not square, but its precise shape is unknown. The formation of 2D bands, interpreted as confinement-induced subbands, is observed experimentally \cite{BenShalom-2010b, Fete-2014, Gariglio-2015b}, but their occupations and energies are uncertain, preventing us from reconstructing a more realistic potential. We do not expect qualitative changes in our conclusions upon going from a square to a triangular or self-consistent confinement potential. The square well, moreover, has the advantage that the pairing matrix elements are known analytically \cite{Valentinis-2016b}. The only unknown parameter of our model is therefore $U$, whose order of magnitude is estimated from first-principles calculations to be a few tens of meV \cite{Stengel-2011}. Figure~\ref{fig:fig2} compares $T_c$ calculated in the square well for various $U$'s with the measured $T_c$ of the 2DEL. The calculation yields critical temperatures similar to those measured and a dome-shaped density dependence for all $U$. This is our main result: localizing STO carriers at a density $\sim\!\!10^{19}$~cm$^{-3}$ into a $\sim\!\!10$-nm-thick layer leads, owing to quantum confinement effects, to a dome in $T_c$ as a function of the sheet carrier density between $1.5$ and $3\times10^{13}$~cm$^{-2}$, with a maximum similar to the maximum $T_c$ of bulk STO. As $L$ becomes large, for $n_{\mathrm{2D}}>2.5\times10^{13}~\mathrm{cm}^{-2}$, the calculated $T_c$ approaches the 3D value (green line) with rapid oscillations for all $U$'s. We propose below tentative explanations for why such oscillations are not seen experimentally. The measured $T_c$ is also very close to the 3D value in this regime, where $T_c$ drops because $n$ actually \emph{decreases} with increasing $n_{\mathrm{2D}}$. In the range where $L\approx10$~nm, the 3D $T_c$ increases monotonically as in Fig.~\ref{fig:fig1}, and one notices a density-dependent critical $U$, above which the $T_c$ at the interface is higher than in the bulk \cite{Valentinis-2016b}. By contrast, in that range the quasi-2D $T_c$ drops after the broad maximum because of the decreasing density-dependent interaction.

\begin{figure}[b]
\includegraphics[width=\columnwidth]{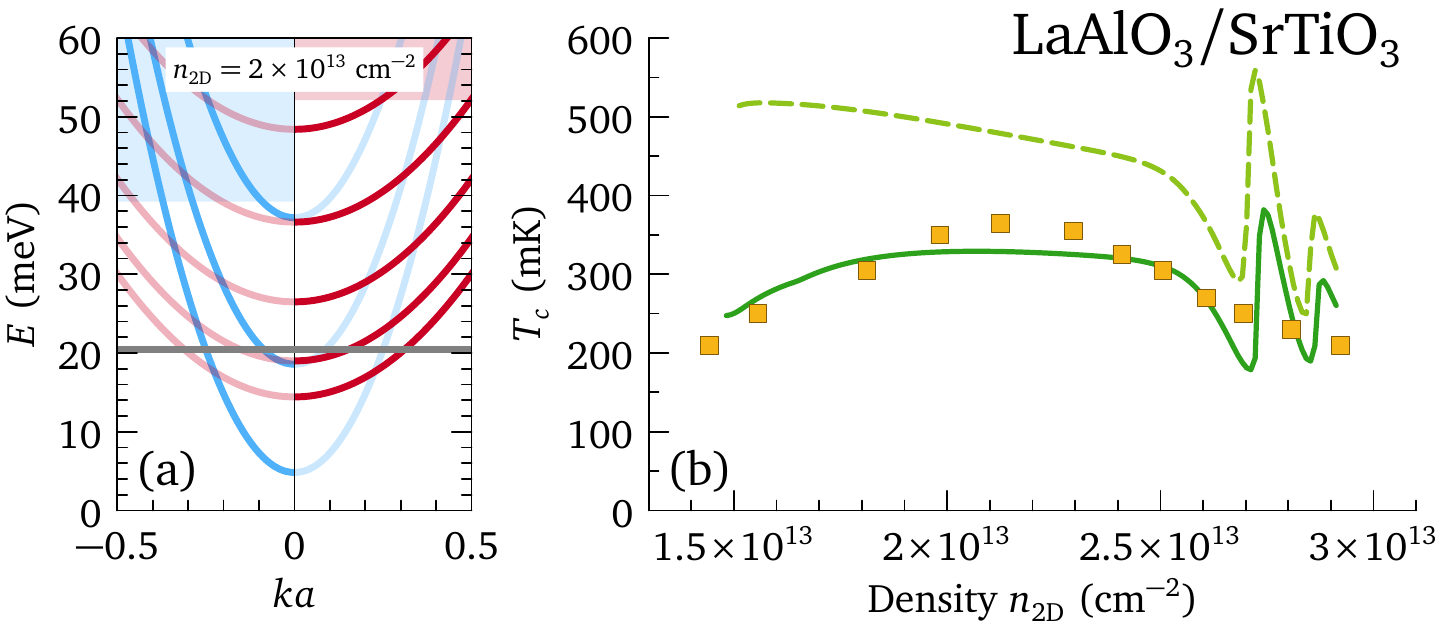}
\caption{\label{fig:fig3}
(a) Subband structure in a quantum well of width 6.8~nm and depth 39~meV for a band splitting $\Delta E=12.9$~meV, corresponding to $n_{\mathrm{2D}}=2\times10^{13}$~cm$^{-2}$. The bound light (heavy) subbands with energies below $U$ ($U+\Delta E$) are shown on the left (right). The chemical potential $\mu=20.5$~meV is also indicated. (b) Measured critical temperature (squares) and calculated $T_c$ (solid line) for band splitting $\Delta E_{\mathrm{STO}}(1+\lambda/\alpha L_{\mathrm{exp}})$ and potential $U=e^2n_{\mathrm{2D}}\alpha L_{\mathrm{exp}}/(\epsilon_0\epsilon)$ with $\alpha=0.65$, $\lambda=30$~nm, and $\epsilon=630$. The dashed line shows $T_c$ calculated with the same $\alpha$, $\lambda=12$~nm, and $\epsilon=377$.
}
\end{figure}

We now refine our model to achieve semiquantitative agreement with the experiment. First, remembering that we mimic a quasitriangular well with a square one, and also because a systematic error in the experimental determination of $L$ cannot be excluded, we replace the measured 2DEL thickness $L_{\mathrm{exp}}$ with $L=\alpha L_{\mathrm{exp}}$ when comparing model and experiment. Second, an increase in the splitting $\Delta E$ is expected at the interface, since both strain and electric field contribute directly to the fourfold crystal field. A modified screening also modifies the Hartree shifts of the bands \cite{Maniv-2015}. We adopt the simple dependence $\Delta E=\Delta E_{\mathrm{STO}}(1+\lambda/L)$ with the fit parameter $\lambda$, which ensures that the bulk STO splitting is recovered for a thick well. Finally, the confinement potential $U$ is in principle linked with $n_{\mathrm{2D}}$ and $L$ via Poisson's equation. The simplest relation results from dimensional analysis as $U=e^2n_{\mathrm{2D}}L/(\epsilon_0\epsilon)$. The model parameters are adjusted to get the best fit displayed in Fig.~\ref{fig:fig3}(b), with $\alpha=0.65$, $\lambda=30$~nm, and $\epsilon=630$. The result $\alpha<1$ is expected, because a square well of width $L_{\mathrm{exp}}$ hosts an electron gas thicker than a triangular well of the same width. Moreover the effective thickness $\alpha L_{\mathrm{exp}}\approx 7$~nm matches the value measured by AFM \cite{Basletic-2008}. Figure~\ref{fig:fig3}(a) shows the subband structure with a splitting five times larger than the bulk splitting at a density of $2\times10^{13}$~cm$^{-2}$. We find that two light and two heavy subbands are partly occupied. Photoemission measurements have reported two light subbands lying below the lowest heavy band \cite{Cancellieri-2014}; our model reproduces this configuration at densities higher than $2.8\times10^{13}$~cm$^{-2}$. The Hall-coefficient nonlinearity observed in Ref.~\onlinecite{Joshua-2012} at $1.6\times10^{13}$~cm$^{-2}$ corresponds in our model to entering the second heavy subband. Finally, the fitted dielectric constant is consistent with the high polarizability of STO: it agrees well with the value $\epsilon=600$ obtained from the low-temperature field-dependent dielectric function of STO, given approximately by $\epsilon=1+\chi_0/(1+\mathscr{E}/\mathscr{E}_0)$ with $\chi_0=24000$ and $\mathscr{E}_0=110$~kV/m \cite{Christen-1994, Copie-2009}, if we substitute for $\mathscr{E}$ the order of magnitude of the interface electric field, e.g., $\mathscr{E}=U/(eL)=4.3$~mV/nm at $n_{\mathrm{2D}}=1.5\times 10^{13}~\mathrm{cm}^{-2}$.

Although we have until now compared the critical temperature with the mean-field $T_c$ from the calculation, in Fig.~\ref{fig:fig3}(b) we show that the model is not inconsistent with the pseudogap scenario suggested by tunneling experiments \cite{Richter-2013}, in which the mean-field $T_c$ continues to increase as the density is reduced, while the critical temperature is suppressed by 2D fluctuations.

\section{Interpretation and discussion}\label{res}

In a thin-film geometry, the critical temperature of a weak-coupling BCS superconductor is a continuous oscillating function of film thickness $L$ and electron density $n$ \cite{Valentinis-2016b}. Pronounced oscillations due to the rapid change in $L$ are visible in Figs.~\ref{fig:fig2} and \ref{fig:fig3}(b), but not resolved in the experiment. Note that the detailed oscillating behavior is sensitive to the shape of the potential well, but the very existence of these oscillations is not, as they simply reflect the presence of subbands. The model involves three types of singularities, which leave specific signatures in $T_c$: a subband lying $\hbar\omega_{\mathrm{D}}$ above the chemical potential begins contributing to $T_c$; the minimum of a subband lies exactly $\hbar\omega_{\mathrm{D}}$ below the chemical potential; and a new subband gets localized in the well. This leads to a quite complex $T_c$ landscape as a function of $L$ and $n$, further complicated by the presence of two types of subbands associated with the bulk light and heavy bands. The $T_c$ maxima do not coincide with any particular singularity of the model but occur approximately when the chemical potential crosses the bottom of a subband. Let us first discuss this rich $T_c$ landscape, before turning to the absence of sharp oscillations in the experiment.

\begin{figure}[tb]
\includegraphics[width=\columnwidth]{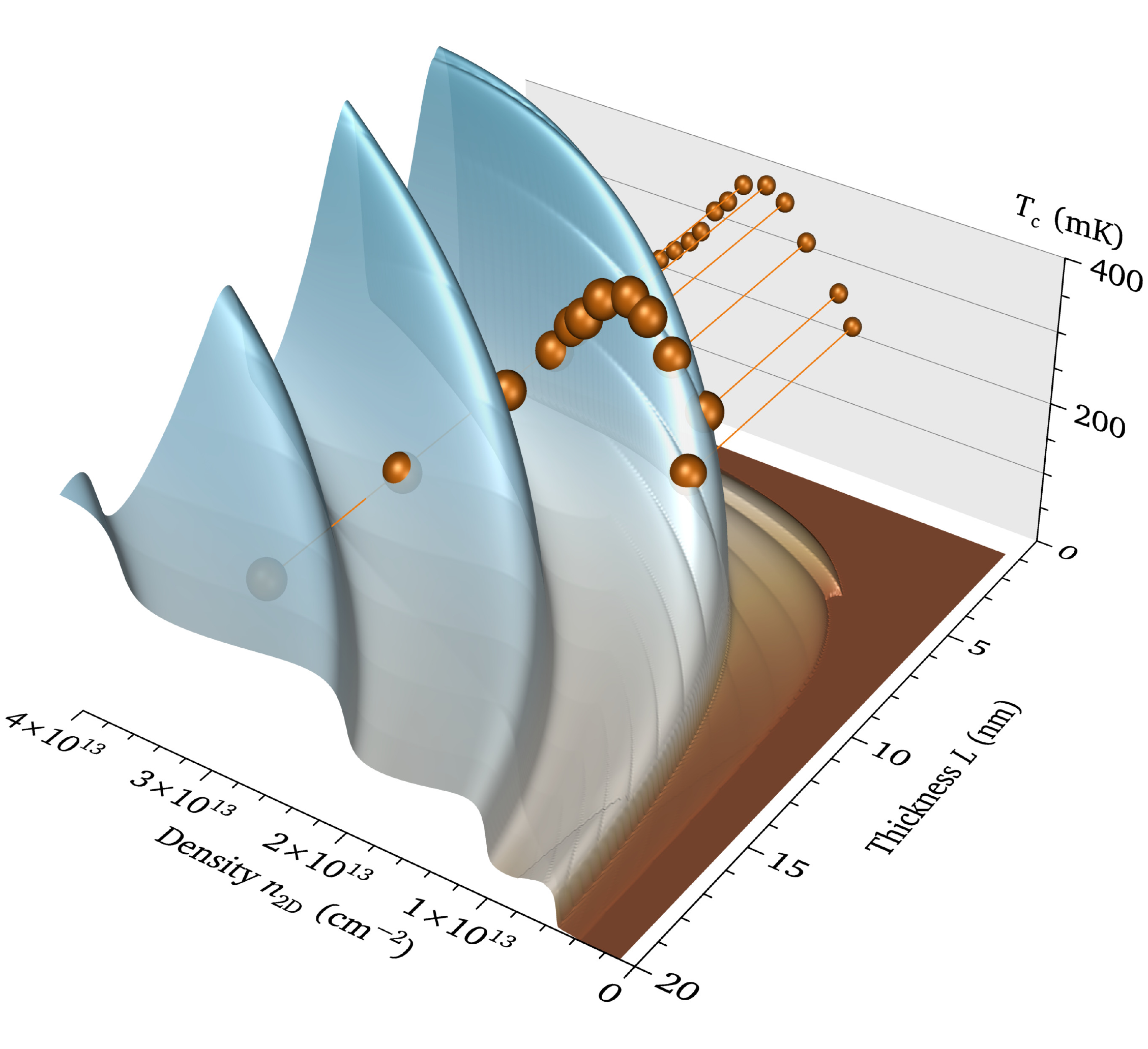}
\caption{\label{fig:fig4}
Critical temperature as a function of thickness $L$ and sheet carrier density $n_{\mathrm{2D}}$. Spheres show the experimental data from Fig.~\ref{fig:fig2}, where the thickness has been reduced by a factor $\alpha=0.65$. The shaded surface shows the $T_c$ calculated with band splitting $\Delta E_{\mathrm{STO}}(1+\lambda/L)$ in a square well of width $L$ and depth $U=e^2n_{\mathrm{2D}}L/(\epsilon_0\epsilon)$, with $\lambda=30$~nm, and $\epsilon=630$.
}
\end{figure}

Using our square-well model, we plot in Fig.~\ref{fig:fig4} the evolution of $T_c$ as a function of both $L$ and $n_{\mathrm{2D}}=nL$ in the domain of thicknesses and densities relevant for the 2DEL. The experimental data are shown for comparison at the rescaled thickness $\alpha L_{\mathrm{exp}}$. To describe the features we start at 5-nm thickness and follow the doping from $0$ upwards. At the lowest doping $T_c$ is exponentially small and reaches a first low-$T_c$ step near $8\times10^{12}~\mathrm{cm}^{-2}$, before rising steeply to $\sim$400~mK for doping of the order of $10^{13}~\mathrm{cm}^{-2}$. At this thickness and for the doping range displayed in the figure, superconductivity is entirely in the two heavy subbands [the two lowest red curves in Fig.~\ref{fig:fig3}(a)]. The low-$T_c$ stage corresponds to the first heavy subband, with a $T_c$ suppressed by a shallow confinement potential (see Fig.~\ref{fig:fig2} for the effect of $U$ on $T_c$) \cite{Valentinis-2016b}. For a fixed 2D carrier density, the number of occupied subbands varies with the layer thickness as $(n_{\mathrm{2D}}L^2)^{1/3}$. Since $T_c$ passes through a maximum when a new heavy subband becomes occupied, the effect of increasing the thickness results in a succession of $T_c$ maxima; the amplitude of these oscillations decays rapidly with increasing thickness since the layer becomes more and more like a 3D bulk sample with a carrier density approaching $0$ according to $n=n_{\mathrm{2D}}/L$. The location of maxima is given approximately by $n_{\mathrm{2D}}\approx\frac{\pi}{12}p(p+1)(4p+5)/L^2$ with integer $p$. The scars running at a small angle with respect to the first maximum are slope changes of $T_c$ occurring when a new heavy subband gets bound in the well as $L$ increases, leading to a slight $T_c$ enhancement. The very faint lines almost parallel to the density axis indicate that a new heavy subband enters the pairing window at $\mu+\hbar\omega_{\mathrm{D}}$. These features reflect that in our calculation we use a BCS-type model interaction with a sharp cutoff at $\hbar\omega_{\mathrm{D}}$. Finally, $\mu$ crossing the third light subband gives another faint structure crossing the previous ones and ending at zero density near $L=20$~nm; other structures associated with the light band are too weak to be visible. Moving up in density for $L=20$~nm, $T_c$ becomes distinguishable from $0$ for $n_{\mathrm{2D}}=3.6\times 10^{12}~\mathrm{cm}^{-2}$, and rises in a succession of steps, each corresponding to a new heavy subband becoming occupied. In the limit of infinite thickness the 3D situation depicted in Fig.~\ref{fig:fig1} is ultimately obtained. The figure illustrates that many combinations of thickness and doping parameters can result in a domelike doping or thickness dependence of $T_c$, however, the maximal value of $T_c$ itself is never strongly different from $T_c^{\max}$ observed in optimally doped bulk STO.

We believe that this theoretical observation provides the main clue as to why $T_c$ is so similar in bulk and interface superconductivity: the pairing potential is basically the same in all cases. Tuning of the density of states by 2D confinement allows the effective pairing interaction to be varied to a certain extent, but its main effect is to define the subband structure. Optimal $T_c$ is found for low densities, which holds true in bulk 3D and quasi-2D alike, thus leading to very similar values of the optimal $T_c$ in 2D and 3D systems.

\begin{figure}[tb]
\includegraphics[width=\columnwidth]{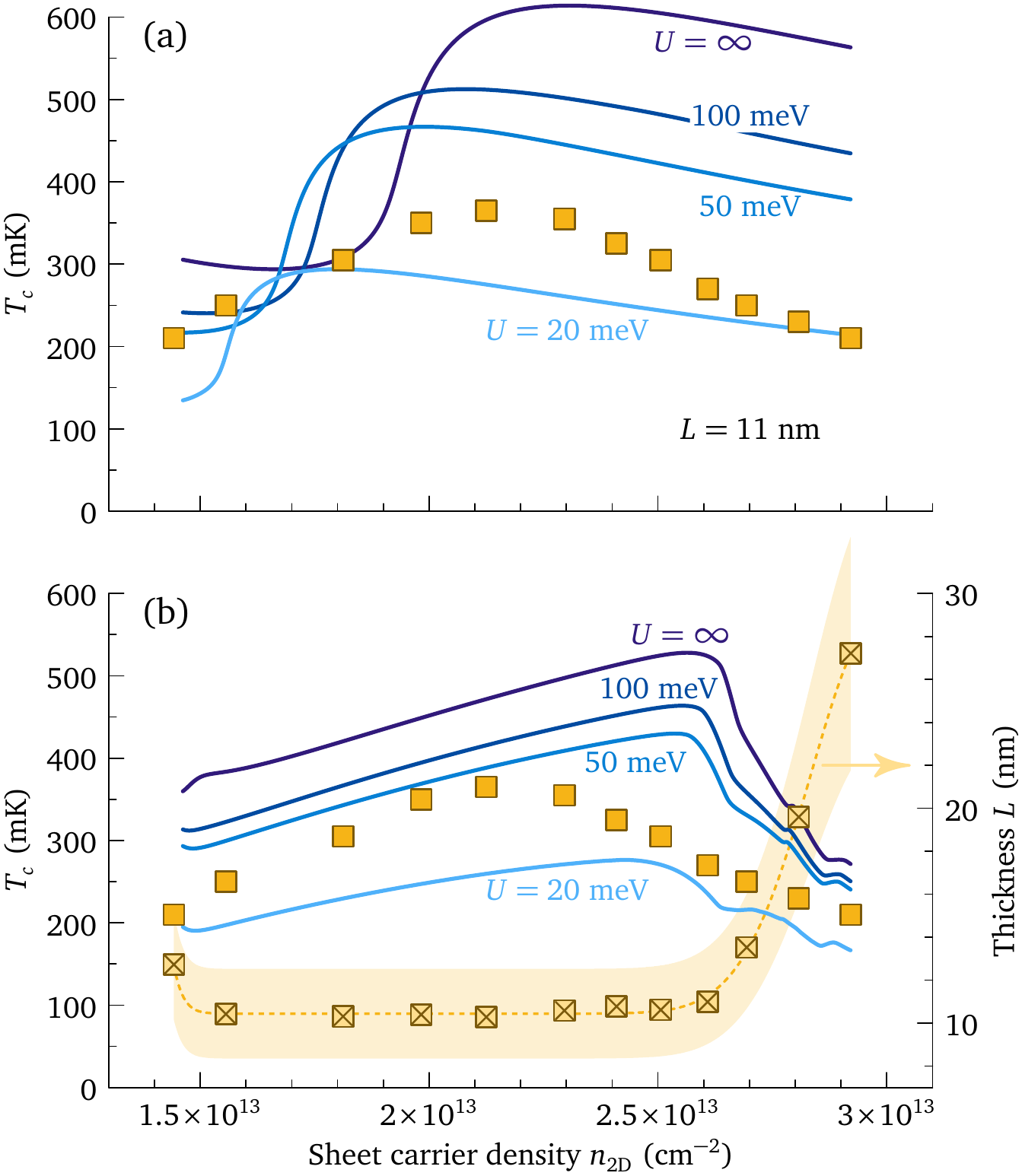}
\caption{\label{fig:fig5}
Same as Fig.~\ref{fig:fig2}, except that (a) the well width is kept fixed to the value $L=11$~nm for all densities; and (b) the well width is varied by $\pm20\%$ around the measured thickness (shaded area) and an average $T_c$ is calculated.
}
\end{figure}

As Fig.~\ref{fig:fig4} clearly demonstrates, the oscillations of $T_c$ for $n_{\mathrm{2D}}>2.5\times 10^{12}~\mathrm{cm}^{-2}$ in Fig.~\ref{fig:fig3}(b) arise due to the abrupt increase in the well thickness used in the calculation, in order to follow the trend indicated by the critical field analysis. It is possible that the variation of the measured superconducting thickness from 10 to 30~nm results less from an actual change in the charge-distribution thickness than from subtle changes in its shape, e.g., the formation of long tails in STO with a very low density. Figure~\ref{fig:fig5}(a) shows that there is no oscillation of the calculated $T_c$ on the overdoped side of the dome if the square-well width stays at a constant value for all densities. We have also examined a scenario where the oscillations are washed out by inhomogeneities. A glimpse at Fig.~\ref{fig:fig4} shows that thickness inhomogeneities are much more effective at suppressing the oscillations than density inhomogeneities. In Fig.~\ref{fig:fig5}(b), we show $T_c$ curves obtained by averaging over a range of thicknesses extending $\pm20\%$ around the measured superconducting thickness, consistent with the estimated experimental uncertainty \cite{Reyren-2009}: this is sufficient to almost completely remove oscillations on the overdoped side. The various simulations displayed in Fig.~\ref{fig:fig5} emphasize that modifications of the model that are able to remove the rapid oscillations of $T_c$ on the overdoped side do not remove the broad maximum between $1.5$ and $3\times10^{13}$~cm$^{-2}$, strengthening our claim that the observed $T_c$ dome may be a shape resonance. In order to gain further insight, one would need to calculate $T_c$ in a triangular or self-consistent quantum well. One difficulty with such a calculation is computing the matrix elements for pairing of the various quantum-well eigenstates---to which $T_c$ is exponentially sensitive---with a sufficient accuracy. The square well is, to our knowledge, the only case where these matrix elements are known analytically \cite{Valentinis-2016b}. The effect of disorder, also expected to smoothen the $T_c$ landscape, should also be investigated.

\section{Conclusion}\label{concl}

In summary, we have explored the hypothesis that the pairing potential responsible for superconductivity is the same in 2D interfaces and 3D bulk \STO{}. We showed that the superconducting dome in the 2D material corresponds to a much narrower effective doping range than in the 3D material, and the optimal $T_c$ in this case coincides with a shape resonance due to the onset of occupation of one of the subbands created by the 3D confining potential. We have shown that the optimal $T_c$ should be very similar in the 2D and 3D cases. A prospect for optimization of $T_c$ is offered by tuning of the confinement potential $U$, which in principle allows $T_c$ to be increased by a significant factor, and by controlled tuning of the thickness parameter $L$ on the 1-nm scale.

\begin{acknowledgments}
This work was supported by the Swiss National Science Foundation through Grants No. 200021-153405 and No. 200021-162628.
\end{acknowledgments}


\begin{thebibliography}{51}%
\makeatletter
\providecommand \@ifxundefined [1]{%
 \@ifx{#1\undefined}
}%
\providecommand \@ifnum [1]{%
 \ifnum #1\expandafter \@firstoftwo
 \else \expandafter \@secondoftwo
 \fi
}%
\providecommand \@ifx [1]{%
 \ifx #1\expandafter \@firstoftwo
 \else \expandafter \@secondoftwo
 \fi
}%
\providecommand \natexlab [1]{#1}%
\providecommand \enquote  [1]{``#1''}%
\providecommand \bibnamefont  [1]{#1}%
\providecommand \bibfnamefont [1]{#1}%
\providecommand \citenamefont [1]{#1}%
\providecommand \href@noop [0]{\@secondoftwo}%
\providecommand \href [0]{\begingroup \@sanitize@url \@href}%
\providecommand \@href[1]{\@@startlink{#1}\@@href}%
\providecommand \@@href[1]{\endgroup#1\@@endlink}%
\providecommand \@sanitize@url [0]{\catcode `\\12\catcode `\$12\catcode
  `\&12\catcode `\#12\catcode `\^12\catcode `\_12\catcode `\%12\relax}%
\providecommand \@@startlink[1]{}%
\providecommand \@@endlink[0]{}%
\providecommand \url  [0]{\begingroup\@sanitize@url \@url }%
\providecommand \@url [1]{\endgroup\@href {#1}{\urlprefix }}%
\providecommand \urlprefix  [0]{URL }%
\providecommand \Eprint [0]{\href }%
\providecommand \doibase [0]{http://dx.doi.org/}%
\providecommand \selectlanguage [0]{\@gobble}%
\providecommand \bibinfo  [0]{\@secondoftwo}%
\providecommand \bibfield  [0]{\@secondoftwo}%
\providecommand \translation [1]{[#1]}%
\providecommand \BibitemOpen [0]{}%
\providecommand \bibitemStop [0]{}%
\providecommand \bibitemNoStop [0]{.\EOS\space}%
\providecommand \EOS [0]{\spacefactor3000\relax}%
\providecommand \BibitemShut  [1]{\csname bibitem#1\endcsname}%
\let\auto@bib@innerbib\@empty
\bibitem [{\citenamefont {Reyren}\ \emph {et~al.}(2007)\citenamefont {Reyren},
  \citenamefont {Thiel}, \citenamefont {Caviglia}, \citenamefont {Kourkoutis},
  \citenamefont {Hammerl}, \citenamefont {Richter}, \citenamefont {Schneider},
  \citenamefont {Kopp}, \citenamefont {R{\"{u}}etschi}, \citenamefont
  {Jaccard}, \citenamefont {Gabay}, \citenamefont {Muller}, \citenamefont
  {Triscone},\ and\ \citenamefont {Mannhart}}]{Reyren-2007}%
  \BibitemOpen
  \bibfield  {author} {\bibinfo {author} {\bibfnamefont {N.}~\bibnamefont
  {Reyren}}, \bibinfo {author} {\bibfnamefont {S.}~\bibnamefont {Thiel}},
  \bibinfo {author} {\bibfnamefont {A.~D.}\ \bibnamefont {Caviglia}}, \bibinfo
  {author} {\bibfnamefont {L.~F.}\ \bibnamefont {Kourkoutis}}, \bibinfo
  {author} {\bibfnamefont {G.}~\bibnamefont {Hammerl}}, \bibinfo {author}
  {\bibfnamefont {C.}~\bibnamefont {Richter}}, \bibinfo {author} {\bibfnamefont
  {C.~W.}\ \bibnamefont {Schneider}}, \bibinfo {author} {\bibfnamefont
  {T.}~\bibnamefont {Kopp}}, \bibinfo {author} {\bibfnamefont {A.-S.}\
  \bibnamefont {R{\"{u}}etschi}}, \bibinfo {author} {\bibfnamefont
  {D.}~\bibnamefont {Jaccard}}, \bibinfo {author} {\bibfnamefont
  {M.}~\bibnamefont {Gabay}}, \bibinfo {author} {\bibfnamefont {D.~A.}\
  \bibnamefont {Muller}}, \bibinfo {author} {\bibfnamefont {J.-M.}\
  \bibnamefont {Triscone}}, \ and\ \bibinfo {author} {\bibfnamefont
  {J.}~\bibnamefont {Mannhart}},\ }\href {\doibase 10.1126/science.1146006}
  {\bibfield  {journal} {\bibinfo  {journal} {Science}\ }\textbf {\bibinfo
  {volume} {317}},\ \bibinfo {pages} {1196} (\bibinfo {year}
  {2007})}\BibitemShut {NoStop}%
\bibitem [{\citenamefont {Klimin}\ \emph {et~al.}(2014)\citenamefont {Klimin},
  \citenamefont {Tempere}, \citenamefont {Devreese},\ and\ \citenamefont
  {van~der Marel}}]{Klimin-2014}%
  \BibitemOpen
  \bibfield  {author} {\bibinfo {author} {\bibfnamefont {S.~N.}\ \bibnamefont
  {Klimin}}, \bibinfo {author} {\bibfnamefont {J.}~\bibnamefont {Tempere}},
  \bibinfo {author} {\bibfnamefont {J.~T.}\ \bibnamefont {Devreese}}, \ and\
  \bibinfo {author} {\bibfnamefont {D.}~\bibnamefont {van~der Marel}},\ }\href
  {\doibase 10.1103/PhysRevB.89.184514} {\bibfield  {journal} {\bibinfo
  {journal} {Phys. Rev. B}\ }\textbf {\bibinfo {volume} {89}},\ \bibinfo
  {pages} {184514} (\bibinfo {year} {2014})}\BibitemShut {NoStop}%
\bibitem [{\citenamefont {Boschker}\ \emph {et~al.}(2015)\citenamefont
  {Boschker}, \citenamefont {Richter}, \citenamefont {Fillis-Tsirakis},
  \citenamefont {Schneider},\ and\ \citenamefont {Mannhart}}]{Boschker-2015}%
  \BibitemOpen
  \bibfield  {author} {\bibinfo {author} {\bibfnamefont {H.}~\bibnamefont
  {Boschker}}, \bibinfo {author} {\bibfnamefont {C.}~\bibnamefont {Richter}},
  \bibinfo {author} {\bibfnamefont {E.}~\bibnamefont {Fillis-Tsirakis}},
  \bibinfo {author} {\bibfnamefont {C.~W.}\ \bibnamefont {Schneider}}, \ and\
  \bibinfo {author} {\bibfnamefont {J.}~\bibnamefont {Mannhart}},\ }\href
  {\doibase 10.1038/srep12309} {\bibfield  {journal} {\bibinfo  {journal} {Sci.
  Rep.}\ }\textbf {\bibinfo {volume} {5}},\ \bibinfo {pages} {12309} (\bibinfo
  {year} {2015})}\BibitemShut {NoStop}%
\bibitem [{\citenamefont {Stephanos}\ \emph {et~al.}(2011)\citenamefont
  {Stephanos}, \citenamefont {Kopp}, \citenamefont {Mannhart},\ and\
  \citenamefont {Hirschfeld}}]{Stephanos-2011}%
  \BibitemOpen
  \bibfield  {author} {\bibinfo {author} {\bibfnamefont {C.}~\bibnamefont
  {Stephanos}}, \bibinfo {author} {\bibfnamefont {T.}~\bibnamefont {Kopp}},
  \bibinfo {author} {\bibfnamefont {J.}~\bibnamefont {Mannhart}}, \ and\
  \bibinfo {author} {\bibfnamefont {P.~J.}\ \bibnamefont {Hirschfeld}},\ }\href
  {\doibase 10.1103/PhysRevB.84.100510} {\bibfield  {journal} {\bibinfo
  {journal} {Phys. Rev. B}\ }\textbf {\bibinfo {volume} {84}},\ \bibinfo
  {pages} {100510} (\bibinfo {year} {2011})}\BibitemShut {NoStop}%
\bibitem [{\citenamefont {Scheurer}\ and\ \citenamefont
  {Schmalian}(2015)}]{Scheurer-2015a}%
  \BibitemOpen
  \bibfield  {author} {\bibinfo {author} {\bibfnamefont {M.~S.}\ \bibnamefont
  {Scheurer}}\ and\ \bibinfo {author} {\bibfnamefont {J.}~\bibnamefont
  {Schmalian}},\ }\href {\doibase doi:10.1038/ncomms7005} {\bibfield  {journal}
  {\bibinfo  {journal} {Nat. Comm.}\ }\textbf {\bibinfo {volume} {6}},\
  \bibinfo {pages} {6005} (\bibinfo {year} {2015})}\BibitemShut {NoStop}%
\bibitem [{\citenamefont {Schooley}\ \emph {et~al.}(1964)\citenamefont
  {Schooley}, \citenamefont {Hosler},\ and\ \citenamefont
  {Cohen}}]{Schooley-1964}%
  \BibitemOpen
  \bibfield  {author} {\bibinfo {author} {\bibfnamefont {J.~F.}\ \bibnamefont
  {Schooley}}, \bibinfo {author} {\bibfnamefont {W.~R.}\ \bibnamefont
  {Hosler}}, \ and\ \bibinfo {author} {\bibfnamefont {M.~L.}\ \bibnamefont
  {Cohen}},\ }\href {\doibase 10.1103/PhysRevLett.12.474} {\bibfield  {journal}
  {\bibinfo  {journal} {Phys. Rev. Lett.}\ }\textbf {\bibinfo {volume} {12}},\
  \bibinfo {pages} {474} (\bibinfo {year} {1964})}\BibitemShut {NoStop}%
\bibitem [{\citenamefont {Schooley}\ \emph {et~al.}(1965)\citenamefont
  {Schooley}, \citenamefont {Hosler}, \citenamefont {Ambler}, \citenamefont
  {Becker}, \citenamefont {Cohen},\ and\ \citenamefont
  {Koonce}}]{Schooley-1965}%
  \BibitemOpen
  \bibfield  {author} {\bibinfo {author} {\bibfnamefont {J.}~\bibnamefont
  {Schooley}}, \bibinfo {author} {\bibfnamefont {W.}~\bibnamefont {Hosler}},
  \bibinfo {author} {\bibfnamefont {E.}~\bibnamefont {Ambler}}, \bibinfo
  {author} {\bibfnamefont {J.}~\bibnamefont {Becker}}, \bibinfo {author}
  {\bibfnamefont {M.}~\bibnamefont {Cohen}}, \ and\ \bibinfo {author}
  {\bibfnamefont {C.}~\bibnamefont {Koonce}},\ }\href 
  {\doibase 10.1103/PhysRevLett.14.305} {\bibfield  {journal} {\bibinfo  {journal} {Phys.
  Rev. Lett.}\ }\textbf {\bibinfo {volume} {14}},\ \bibinfo {pages} {305}
  (\bibinfo {year} {1965})}\BibitemShut {NoStop}%
\bibitem [{\citenamefont {Koonce}\ \emph {et~al.}(1967)\citenamefont {Koonce},
  \citenamefont {Cohen}, \citenamefont {Schooley}, \citenamefont {Hosler},\
  and\ \citenamefont {Pfeiffer}}]{Koonce-1967}%
  \BibitemOpen
  \bibfield  {author} {\bibinfo {author} {\bibfnamefont {C.~S.}\ \bibnamefont
  {Koonce}}, \bibinfo {author} {\bibfnamefont {M.~L.}\ \bibnamefont {Cohen}},
  \bibinfo {author} {\bibfnamefont {J.~F.}\ \bibnamefont {Schooley}}, \bibinfo
  {author} {\bibfnamefont {W.~R.}\ \bibnamefont {Hosler}}, \ and\ \bibinfo
  {author} {\bibfnamefont {E.~R.}\ \bibnamefont {Pfeiffer}},\ }\href 
  {\doibase 10.1103/PhysRev.163.380} {\bibfield  {journal} {\bibinfo  {journal} {Phys.
  Rev.}\ }\textbf {\bibinfo {volume} {163}},\ \bibinfo {pages} {380} (\bibinfo
  {year} {1967})}\BibitemShut {NoStop}%
\bibitem [{\citenamefont {Lin}\ \emph {et~al.}(2014)\citenamefont {Lin},
  \citenamefont {Bridoux}, \citenamefont {Gourgout}, \citenamefont {Seyfarth},
  \citenamefont {Kr{\"a}mer}, \citenamefont {Nardone}, \citenamefont
  {Fauqu{\'e}},\ and\ \citenamefont {Behnia}}]{Lin-2014}%
  \BibitemOpen
  \bibfield  {author} {\bibinfo {author} {\bibfnamefont {X.}~\bibnamefont
  {Lin}}, \bibinfo {author} {\bibfnamefont {G.}~\bibnamefont {Bridoux}},
  \bibinfo {author} {\bibfnamefont {A.}~\bibnamefont {Gourgout}}, \bibinfo
  {author} {\bibfnamefont {G.}~\bibnamefont {Seyfarth}}, \bibinfo {author}
  {\bibfnamefont {S.}~\bibnamefont {Kr{\"a}mer}}, \bibinfo {author}
  {\bibfnamefont {M.}~\bibnamefont {Nardone}}, \bibinfo {author} {\bibfnamefont
  {B.}~\bibnamefont {Fauqu{\'e}}}, \ and\ \bibinfo {author} {\bibfnamefont
  {K.}~\bibnamefont {Behnia}},\ }\href 
  {\doibase 10.1103/PhysRevLett.112.207002} {\bibfield  {journal} {\bibinfo  {journal}
  {Phys. Rev. Lett.}\ }\textbf {\bibinfo {volume} {112}},\ \bibinfo {pages}
  {207002} (\bibinfo {year} {2014})}\BibitemShut {NoStop}%
\bibitem [{\citenamefont {Caviglia}\ \emph {et~al.}(2008)\citenamefont
  {Caviglia}, \citenamefont {Gariglio}, \citenamefont {Reyren}, \citenamefont
  {Jaccard}, \citenamefont {Schneider}, \citenamefont {Gabay}, \citenamefont
  {Thiel}, \citenamefont {Hammerl}, \citenamefont {Mannhart},\ and\
  \citenamefont {Triscone}}]{Caviglia-2008}%
  \BibitemOpen
  \bibfield  {author} {\bibinfo {author} {\bibfnamefont {A.~D.}\ \bibnamefont
  {Caviglia}}, \bibinfo {author} {\bibfnamefont {S.}~\bibnamefont {Gariglio}},
  \bibinfo {author} {\bibfnamefont {N.}~\bibnamefont {Reyren}}, \bibinfo
  {author} {\bibfnamefont {D.}~\bibnamefont {Jaccard}}, \bibinfo {author}
  {\bibfnamefont {T.}~\bibnamefont {Schneider}}, \bibinfo {author}
  {\bibfnamefont {M.}~\bibnamefont {Gabay}}, \bibinfo {author} {\bibfnamefont
  {S.}~\bibnamefont {Thiel}}, \bibinfo {author} {\bibfnamefont
  {G.}~\bibnamefont {Hammerl}}, \bibinfo {author} {\bibfnamefont
  {J.}~\bibnamefont {Mannhart}}, \ and\ \bibinfo {author} {\bibfnamefont
  {J.-M.}\ \bibnamefont {Triscone}},\ }\href {\doibase 10.1038/nature07576}
  {\bibfield  {journal} {\bibinfo  {journal} {Nature}\ }\textbf {\bibinfo
  {volume} {456}},\ \bibinfo {pages} {624} (\bibinfo {year}
  {2008})}\BibitemShut {NoStop}%
\bibitem [{\citenamefont {{Joshua}}\ \emph {et~al.}(2012)\citenamefont
  {{Joshua}}, \citenamefont {{Pecker}}, \citenamefont {{Ruhman}}, \citenamefont
  {{Altman}},\ and\ \citenamefont {{Ilani}}}]{Joshua-2012}%
  \BibitemOpen
  \bibfield  {author} {\bibinfo {author} {\bibfnamefont {A.}~\bibnamefont
  {{Joshua}}}, \bibinfo {author} {\bibfnamefont {S.}~\bibnamefont {{Pecker}}},
  \bibinfo {author} {\bibfnamefont {J.}~\bibnamefont {{Ruhman}}}, \bibinfo
  {author} {\bibfnamefont {E.}~\bibnamefont {{Altman}}}, \ and\ \bibinfo
  {author} {\bibfnamefont {S.}~\bibnamefont {{Ilani}}},\ }\href 
  {\doibase 10.1038/ncomms2116} {\bibfield  {journal} {\bibinfo  {journal} {Nat. Comm.}\
  }\textbf {\bibinfo {volume} {3}},\ \bibinfo {eid} {1129} (\bibinfo {year}
  {2012})}\BibitemShut {NoStop}%
\bibitem [{\citenamefont {Bell}\ \emph {et~al.}(2009)\citenamefont {Bell},
  \citenamefont {Harashima}, \citenamefont {Kozuka}, \citenamefont {Kim},
  \citenamefont {Kim}, \citenamefont {Hikita},\ and\ \citenamefont
  {Hwang}}]{Bell-2009}%
  \BibitemOpen
  \bibfield  {author} {\bibinfo {author} {\bibfnamefont {C.}~\bibnamefont
  {Bell}}, \bibinfo {author} {\bibfnamefont {S.}~\bibnamefont {Harashima}},
  \bibinfo {author} {\bibfnamefont {Y.}~\bibnamefont {Kozuka}}, \bibinfo
  {author} {\bibfnamefont {M.}~\bibnamefont {Kim}}, \bibinfo {author}
  {\bibfnamefont {B.~G.}\ \bibnamefont {Kim}}, \bibinfo {author} {\bibfnamefont
  {Y.}~\bibnamefont {Hikita}}, \ and\ \bibinfo {author} {\bibfnamefont {H.~Y.}\
  \bibnamefont {Hwang}},\ }\href {\doibase 10.1103/PhysRevLett.103.226802}
  {\bibfield  {journal} {\bibinfo  {journal} {Phys. Rev. Lett.}\ }\textbf
  {\bibinfo {volume} {103}},\ \bibinfo {pages} {226802} (\bibinfo {year}
  {2009})}\BibitemShut {NoStop}%
\bibitem [{\citenamefont {Bert}\ \emph {et~al.}(2012)\citenamefont {Bert},
  \citenamefont {Nowack}, \citenamefont {Kalisky}, \citenamefont {Noad},
  \citenamefont {Kirtley}, \citenamefont {Bell}, \citenamefont {Sato},
  \citenamefont {Hosoda}, \citenamefont {Hikita}, \citenamefont {Hwang},\ and\
  \citenamefont {Moler}}]{Bert-2012}%
  \BibitemOpen
  \bibfield  {author} {\bibinfo {author} {\bibfnamefont {J.~A.}\ \bibnamefont
  {Bert}}, \bibinfo {author} {\bibfnamefont {K.~C.}\ \bibnamefont {Nowack}},
  \bibinfo {author} {\bibfnamefont {B.}~\bibnamefont {Kalisky}}, \bibinfo
  {author} {\bibfnamefont {H.}~\bibnamefont {Noad}}, \bibinfo {author}
  {\bibfnamefont {J.~R.}\ \bibnamefont {Kirtley}}, \bibinfo {author}
  {\bibfnamefont {C.}~\bibnamefont {Bell}}, \bibinfo {author} {\bibfnamefont
  {H.~K.}\ \bibnamefont {Sato}}, \bibinfo {author} {\bibfnamefont
  {M.}~\bibnamefont {Hosoda}}, \bibinfo {author} {\bibfnamefont
  {Y.}~\bibnamefont {Hikita}}, \bibinfo {author} {\bibfnamefont {H.~Y.}\
  \bibnamefont {Hwang}}, \ and\ \bibinfo {author} {\bibfnamefont {K.~A.}\
  \bibnamefont {Moler}},\ }\href {\doibase 10.1103/PhysRevB.86.060503}
  {\bibfield  {journal} {\bibinfo  {journal} {Phys. Rev. B}\ }\textbf {\bibinfo
  {volume} {86}},\ \bibinfo {pages} {060503} (\bibinfo {year}
  {2012})}\BibitemShut {NoStop}%
\bibitem [{\citenamefont {Copie}\ \emph {et~al.}(2009)\citenamefont {Copie},
  \citenamefont {Garcia}, \citenamefont {B{\"{o}}defeld}, \citenamefont
  {Carr{\'{e}}t{\'{e}}ro}, \citenamefont {Bibes}, \citenamefont {Herranz},
  \citenamefont {Jacquet}, \citenamefont {Maurice}, \citenamefont {Vinter},
  \citenamefont {Fusil}, \citenamefont {Bouzehouane}, \citenamefont
  {Jaffr{\`{e}}s},\ and\ \citenamefont {Barth{\'{e}}l{\'{e}}my}}]{Copie-2009}%
  \BibitemOpen
  \bibfield  {author} {\bibinfo {author} {\bibfnamefont {O.}~\bibnamefont
  {Copie}}, \bibinfo {author} {\bibfnamefont {V.}~\bibnamefont {Garcia}},
  \bibinfo {author} {\bibfnamefont {C.}~\bibnamefont {B{\"{o}}defeld}},
  \bibinfo {author} {\bibfnamefont {C.}~\bibnamefont {Carr{\'{e}}t{\'{e}}ro}},
  \bibinfo {author} {\bibfnamefont {M.}~\bibnamefont {Bibes}}, \bibinfo
  {author} {\bibfnamefont {G.}~\bibnamefont {Herranz}}, \bibinfo {author}
  {\bibfnamefont {E.}~\bibnamefont {Jacquet}}, \bibinfo {author} {\bibfnamefont
  {J.-L.}\ \bibnamefont {Maurice}}, \bibinfo {author} {\bibfnamefont
  {B.}~\bibnamefont {Vinter}}, \bibinfo {author} {\bibfnamefont
  {S.}~\bibnamefont {Fusil}}, \bibinfo {author} {\bibfnamefont
  {K.}~\bibnamefont {Bouzehouane}}, \bibinfo {author} {\bibfnamefont
  {H.}~\bibnamefont {Jaffr{\`{e}}s}}, \ and\ \bibinfo {author} {\bibfnamefont
  {A.}~\bibnamefont {Barth{\'{e}}l{\'{e}}my}},\ }\href 
  {\doibase 10.1103/PhysRevLett.102.216804} {\bibfield  {journal} {\bibinfo  {journal}
  {Phys. Rev. Lett.}\ }\textbf {\bibinfo {volume} {102}},\ \bibinfo {pages}
  {216804} (\bibinfo {year} {2009})}\BibitemShut {NoStop}%
\bibitem [{\citenamefont {Dubroka}\ \emph {et~al.}(2010)\citenamefont
  {Dubroka}, \citenamefont {R{\"o}ssle}, \citenamefont {Kim}, \citenamefont
  {Malik}, \citenamefont {Schultz}, \citenamefont {Thiel}, \citenamefont
  {Schneider}, \citenamefont {Mannhart}, \citenamefont {Herranz}, \citenamefont
  {Copie}, \citenamefont {Bibes}, \citenamefont {Barth{\'e}l{\'e}my},\ and\
  \citenamefont {Bernhard}}]{Dubroka-2010}%
  \BibitemOpen
  \bibfield  {author} {\bibinfo {author} {\bibfnamefont {A.}~\bibnamefont
  {Dubroka}}, \bibinfo {author} {\bibfnamefont {M.}~\bibnamefont {R{\"o}ssle}},
  \bibinfo {author} {\bibfnamefont {K.~W.}\ \bibnamefont {Kim}}, \bibinfo
  {author} {\bibfnamefont {V.~K.}\ \bibnamefont {Malik}}, \bibinfo {author}
  {\bibfnamefont {L.}~\bibnamefont {Schultz}}, \bibinfo {author} {\bibfnamefont
  {S.}~\bibnamefont {Thiel}}, \bibinfo {author} {\bibfnamefont {C.~W.}\
  \bibnamefont {Schneider}}, \bibinfo {author} {\bibfnamefont {J.}~\bibnamefont
  {Mannhart}}, \bibinfo {author} {\bibfnamefont {G.}~\bibnamefont {Herranz}},
  \bibinfo {author} {\bibfnamefont {O.}~\bibnamefont {Copie}}, \bibinfo
  {author} {\bibfnamefont {M.}~\bibnamefont {Bibes}}, \bibinfo {author}
  {\bibfnamefont {A.}~\bibnamefont {Barth{\'e}l{\'e}my}}, \ and\ \bibinfo
  {author} {\bibfnamefont {C.}~\bibnamefont {Bernhard}},\ }\href 
  {\doibase 10.1103/PhysRevLett.104.156807} {\bibfield  {journal} {\bibinfo  {journal}
  {Phys. Rev. Lett.}\ }\textbf {\bibinfo {volume} {104}},\ \bibinfo {pages}
  {156807} (\bibinfo {year} {2010})}\BibitemShut {NoStop}%
\bibitem [{\citenamefont {Gariglio}\ \emph
  {et~al.}(2015{\natexlab{a}})\citenamefont {Gariglio}, \citenamefont
  {F{\^e}te},\ and\ \citenamefont {Triscone}}]{Gariglio-2015b}%
  \BibitemOpen
  \bibfield  {author} {\bibinfo {author} {\bibfnamefont {S.}~\bibnamefont
  {Gariglio}}, \bibinfo {author} {\bibfnamefont {A.}~\bibnamefont {F{\^e}te}},
  \ and\ \bibinfo {author} {\bibfnamefont {J.-M.}\ \bibnamefont {Triscone}},\
  }\href {\doibase 10.1088/0953-8984/27/28/283201} {\bibfield  {journal}
  {\bibinfo  {journal} {J. Phys.: Condens. Mat.}\ }\textbf {\bibinfo {volume}
  {27}},\ \bibinfo {pages} {283201} (\bibinfo {year}
  {2015}{\natexlab{a}})}\BibitemShut {NoStop}%
\bibitem [{\citenamefont {Maniv}\ \emph {et~al.}(2015)\citenamefont {Maniv},
  \citenamefont {Ben~Shalom}, \citenamefont {Ron}, \citenamefont {Mograbi},
  \citenamefont {Palevski}, \citenamefont {Goldstein},\ and\ \citenamefont
  {Dagan}}]{Maniv-2015}%
  \BibitemOpen
  \bibfield  {author} {\bibinfo {author} {\bibfnamefont {E.}~\bibnamefont
  {Maniv}}, \bibinfo {author} {\bibfnamefont {M.}~\bibnamefont {Ben~Shalom}},
  \bibinfo {author} {\bibfnamefont {A.}~\bibnamefont {Ron}}, \bibinfo {author}
  {\bibfnamefont {M.}~\bibnamefont {Mograbi}}, \bibinfo {author} {\bibfnamefont
  {A.}~\bibnamefont {Palevski}}, \bibinfo {author} {\bibfnamefont
  {M.}~\bibnamefont {Goldstein}}, \ and\ \bibinfo {author} {\bibfnamefont
  {Y.}~\bibnamefont {Dagan}},\ }\href {\doibase 10.1038/ncomms9239} {\bibfield
  {journal} {\bibinfo  {journal} {Nat. Comm.}\ }\textbf {\bibinfo {volume}
  {6}},\ \bibinfo {pages} {8239} (\bibinfo {year} {2015})}\BibitemShut
  {NoStop}%
\bibitem [{\citenamefont {Gor'kov}\ and\ \citenamefont
  {Rashba}(2001)}]{Gorkov-2001}%
  \BibitemOpen
  \bibfield  {author} {\bibinfo {author} {\bibfnamefont {L.~P.}\ \bibnamefont
  {Gor'kov}}\ and\ \bibinfo {author} {\bibfnamefont {E.~I.}\ \bibnamefont
  {Rashba}},\ }\href {\doibase 10.1103/PhysRevLett.87.037004} {\bibfield
  {journal} {\bibinfo  {journal} {Phys. Rev. Lett.}\ }\textbf {\bibinfo
  {volume} {87}},\ \bibinfo {pages} {037004} (\bibinfo {year}
  {2001})}\BibitemShut {NoStop}%
\bibitem [{\citenamefont {Kirtley}\ \emph {et~al.}(2012)\citenamefont
  {Kirtley}, \citenamefont {Kalisky}, \citenamefont {Bert}, \citenamefont
  {Bell}, \citenamefont {Kim}, \citenamefont {Hikita}, \citenamefont {Hwang},
  \citenamefont {Ngai}, \citenamefont {Segal}, \citenamefont {Walker},
  \citenamefont {Ahn},\ and\ \citenamefont {Moler}}]{Kirtley-2012}%
  \BibitemOpen
  \bibfield  {author} {\bibinfo {author} {\bibfnamefont {J.~R.}\ \bibnamefont
  {Kirtley}}, \bibinfo {author} {\bibfnamefont {B.}~\bibnamefont {Kalisky}},
  \bibinfo {author} {\bibfnamefont {J.~A.}\ \bibnamefont {Bert}}, \bibinfo
  {author} {\bibfnamefont {C.}~\bibnamefont {Bell}}, \bibinfo {author}
  {\bibfnamefont {M.}~\bibnamefont {Kim}}, \bibinfo {author} {\bibfnamefont
  {Y.}~\bibnamefont {Hikita}}, \bibinfo {author} {\bibfnamefont {H.~Y.}\
  \bibnamefont {Hwang}}, \bibinfo {author} {\bibfnamefont {J.~H.}\ \bibnamefont
  {Ngai}}, \bibinfo {author} {\bibfnamefont {Y.}~\bibnamefont {Segal}},
  \bibinfo {author} {\bibfnamefont {F.~J.}\ \bibnamefont {Walker}}, \bibinfo
  {author} {\bibfnamefont {C.~H.}\ \bibnamefont {Ahn}}, \ and\ \bibinfo
  {author} {\bibfnamefont {K.~A.}\ \bibnamefont {Moler}},\ }\href 
  {\doibase 10.1103/PhysRevB.85.224518} {\bibfield  {journal} {\bibinfo  {journal} {Phys.
  Rev. B}\ }\textbf {\bibinfo {volume} {85}},\ \bibinfo {pages} {224518}
  (\bibinfo {year} {2012})}\BibitemShut {NoStop}%
\bibitem [{\citenamefont {Benfatto}\ \emph {et~al.}(2009)\citenamefont
  {Benfatto}, \citenamefont {Castellani},\ and\ \citenamefont
  {Giamarchi}}]{Benfatto-2009}%
  \BibitemOpen
  \bibfield  {author} {\bibinfo {author} {\bibfnamefont {L.}~\bibnamefont
  {Benfatto}}, \bibinfo {author} {\bibfnamefont {C.}~\bibnamefont
  {Castellani}}, \ and\ \bibinfo {author} {\bibfnamefont {T.}~\bibnamefont
  {Giamarchi}},\ }\href {\doibase 10.1103/PhysRevB.80.214506} {\bibfield
  {journal} {\bibinfo  {journal} {Phys. Rev. B}\ }\textbf {\bibinfo {volume}
  {80}},\ \bibinfo {pages} {214506} (\bibinfo {year} {2009})}\BibitemShut
  {NoStop}%
\bibitem [{\citenamefont {Li}\ \emph {et~al.}(2011)\citenamefont {Li},
  \citenamefont {Richter}, \citenamefont {Mannhart},\ and\ \citenamefont
  {Ashoori}}]{Li-2011}%
  \BibitemOpen
  \bibfield  {author} {\bibinfo {author} {\bibfnamefont {L.}~\bibnamefont
  {Li}}, \bibinfo {author} {\bibfnamefont {C.}~\bibnamefont {Richter}},
  \bibinfo {author} {\bibfnamefont {J.}~\bibnamefont {Mannhart}}, \ and\
  \bibinfo {author} {\bibfnamefont {R.}~\bibnamefont {Ashoori}},\ }\href
  {http://www.nature.com/nphys/journal/v7/n10/full/nphys2080.html} {\bibfield
  {journal} {\bibinfo  {journal} {Nature Physics}\ }\textbf {\bibinfo {volume}
  {7}},\ \bibinfo {pages} {762} (\bibinfo {year} {2011})}\BibitemShut {NoStop}%
\bibitem [{\citenamefont {Bert}\ \emph {et~al.}(2011)\citenamefont {Bert},
  \citenamefont {Kalisky}, \citenamefont {Bell}, \citenamefont {Kim},
  \citenamefont {Hikita}, \citenamefont {Hwang},\ and\ \citenamefont
  {Moler}}]{Bert-2011}%
  \BibitemOpen
  \bibfield  {author} {\bibinfo {author} {\bibfnamefont {J.~A.}\ \bibnamefont
  {Bert}}, \bibinfo {author} {\bibfnamefont {B.}~\bibnamefont {Kalisky}},
  \bibinfo {author} {\bibfnamefont {C.}~\bibnamefont {Bell}}, \bibinfo {author}
  {\bibfnamefont {M.}~\bibnamefont {Kim}}, \bibinfo {author} {\bibfnamefont
  {Y.}~\bibnamefont {Hikita}}, \bibinfo {author} {\bibfnamefont {H.~Y.}\
  \bibnamefont {Hwang}}, \ and\ \bibinfo {author} {\bibfnamefont {K.~A.}\
  \bibnamefont {Moler}},\ }\href
  {http://www.nature.com/nphys/journal/v7/n10/full/nphys2079.html} {\bibfield
  {journal} {\bibinfo  {journal} {Nature Physics}\ }\textbf {\bibinfo {volume}
  {7}},\ \bibinfo {pages} {767} (\bibinfo {year} {2011})}\BibitemShut {NoStop}%
\bibitem [{\citenamefont {Dikin}\ \emph {et~al.}(2011)\citenamefont {Dikin},
  \citenamefont {Mehta}, \citenamefont {Bark}, \citenamefont {Folkman},
  \citenamefont {Eom},\ and\ \citenamefont {Chandrasekhar}}]{Dikin-2011}%
  \BibitemOpen
  \bibfield  {author} {\bibinfo {author} {\bibfnamefont {D.~A.}\ \bibnamefont
  {Dikin}}, \bibinfo {author} {\bibfnamefont {M.}~\bibnamefont {Mehta}},
  \bibinfo {author} {\bibfnamefont {C.~W.}\ \bibnamefont {Bark}}, \bibinfo
  {author} {\bibfnamefont {C.~M.}\ \bibnamefont {Folkman}}, \bibinfo {author}
  {\bibfnamefont {C.~B.}\ \bibnamefont {Eom}}, \ and\ \bibinfo {author}
  {\bibfnamefont {V.}~\bibnamefont {Chandrasekhar}},\ }\href 
  {\doibase 10.1103/PhysRevLett.107.056802} {\bibfield  {journal} {\bibinfo  {journal}
  {Phys. Rev. Lett.}\ }\textbf {\bibinfo {volume} {107}},\ \bibinfo {pages}
  {056802} (\bibinfo {year} {2011})}\BibitemShut {NoStop}%
\bibitem [{\citenamefont {Daptary}\ \emph {et~al.}(2017)\citenamefont
  {Daptary}, \citenamefont {Kumar}, \citenamefont {Bid}, \citenamefont {Kumar},
  \citenamefont {Dogra}, \citenamefont {Budhani}, \citenamefont {Kumar},
  \citenamefont {Mohanta},\ and\ \citenamefont {Taraphder}}]{Daptary-2017}%
  \BibitemOpen
  \bibfield  {author} {\bibinfo {author} {\bibfnamefont {G.~N.}\ \bibnamefont
  {Daptary}}, \bibinfo {author} {\bibfnamefont {S.}~\bibnamefont {Kumar}},
  \bibinfo {author} {\bibfnamefont {A.}~\bibnamefont {Bid}}, \bibinfo {author}
  {\bibfnamefont {P.}~\bibnamefont {Kumar}}, \bibinfo {author} {\bibfnamefont
  {A.}~\bibnamefont {Dogra}}, \bibinfo {author} {\bibfnamefont {R.~C.}\
  \bibnamefont {Budhani}}, \bibinfo {author} {\bibfnamefont {D.}~\bibnamefont
  {Kumar}}, \bibinfo {author} {\bibfnamefont {N.}~\bibnamefont {Mohanta}}, \
  and\ \bibinfo {author} {\bibfnamefont {A.}~\bibnamefont {Taraphder}},\ }\href
  {\doibase 10.1103/PhysRevB.95.174502} {\bibfield  {journal} {\bibinfo
  {journal} {Phys. Rev. B}\ }\textbf {\bibinfo {volume} {95}},\ \bibinfo
  {pages} {174502} (\bibinfo {year} {2017})}\BibitemShut {NoStop}%
\bibitem [{\citenamefont {Caprara}\ \emph {et~al.}(2012)\citenamefont
  {Caprara}, \citenamefont {Peronaci},\ and\ \citenamefont
  {Grilli}}]{Caprara-2012}%
  \BibitemOpen
  \bibfield  {author} {\bibinfo {author} {\bibfnamefont {S.}~\bibnamefont
  {Caprara}}, \bibinfo {author} {\bibfnamefont {F.}~\bibnamefont {Peronaci}}, \
  and\ \bibinfo {author} {\bibfnamefont {M.}~\bibnamefont {Grilli}},\ }\href
  {\doibase 10.1103/PhysRevLett.109.196401} {\bibfield  {journal} {\bibinfo
  {journal} {Phys. Rev. Lett.}\ }\textbf {\bibinfo {volume} {109}},\ \bibinfo
  {pages} {196401} (\bibinfo {year} {2012})}\BibitemShut {NoStop}%
\bibitem [{\citenamefont {Caprara}\ \emph {et~al.}(2013)\citenamefont
  {Caprara}, \citenamefont {Biscaras}, \citenamefont {Bergeal}, \citenamefont
  {Bucheli}, \citenamefont {Hurand}, \citenamefont {Feuillet-Palma},
  \citenamefont {Rastogi}, \citenamefont {Budhani}, \citenamefont {Lesueur},\
  and\ \citenamefont {Grilli}}]{Caprara-2013}%
  \BibitemOpen
  \bibfield  {author} {\bibinfo {author} {\bibfnamefont {S.}~\bibnamefont
  {Caprara}}, \bibinfo {author} {\bibfnamefont {J.}~\bibnamefont {Biscaras}},
  \bibinfo {author} {\bibfnamefont {N.}~\bibnamefont {Bergeal}}, \bibinfo
  {author} {\bibfnamefont {D.}~\bibnamefont {Bucheli}}, \bibinfo {author}
  {\bibfnamefont {S.}~\bibnamefont {Hurand}}, \bibinfo {author} {\bibfnamefont
  {C.}~\bibnamefont {Feuillet-Palma}}, \bibinfo {author} {\bibfnamefont
  {A.}~\bibnamefont {Rastogi}}, \bibinfo {author} {\bibfnamefont {R.~C.}\
  \bibnamefont {Budhani}}, \bibinfo {author} {\bibfnamefont {J.}~\bibnamefont
  {Lesueur}}, \ and\ \bibinfo {author} {\bibfnamefont {M.}~\bibnamefont
  {Grilli}},\ }\href {\doibase 10.1103/PhysRevB.88.020504} {\bibfield
  {journal} {\bibinfo  {journal} {Phys. Rev. B}\ }\textbf {\bibinfo {volume}
  {88}},\ \bibinfo {pages} {020504} (\bibinfo {year} {2013})}\BibitemShut
  {NoStop}%
\bibitem [{\citenamefont {Bianconi}\ \emph {et~al.}(2014)\citenamefont
  {Bianconi}, \citenamefont {Innocenti}, \citenamefont {Valletta},\ and\
  \citenamefont {Perali}}]{Bianconi-2014}%
  \BibitemOpen
  \bibfield  {author} {\bibinfo {author} {\bibfnamefont {A.}~\bibnamefont
  {Bianconi}}, \bibinfo {author} {\bibfnamefont {D.}~\bibnamefont {Innocenti}},
  \bibinfo {author} {\bibfnamefont {A.}~\bibnamefont {Valletta}}, \ and\
  \bibinfo {author} {\bibfnamefont {A.}~\bibnamefont {Perali}},\ }\href
  {\doibase 10.1088/1742-6596/529/1/012007} {\bibfield  {journal} {\bibinfo
  {journal} {J. Phys.: Conf. Ser.}\ }\textbf {\bibinfo {volume} {529}},\
  \bibinfo {pages} {012007} (\bibinfo {year} {2014})}\BibitemShut {NoStop}%
\bibitem [{\citenamefont {Thompson}\ and\ \citenamefont
  {Blatt}(1963)}]{Thompson-1963}%
  \BibitemOpen
  \bibfield  {author} {\bibinfo {author} {\bibfnamefont {C.~J.}\ \bibnamefont
  {Thompson}}\ and\ \bibinfo {author} {\bibfnamefont {J.~M.}\ \bibnamefont
  {Blatt}},\ }\href {\doibase 10.1016/S0375-9601(63)80003-1} {\bibfield
  {journal} {\bibinfo  {journal} {Phys. Lett.}\ }\textbf {\bibinfo {volume}
  {5}},\ \bibinfo {pages} {6} (\bibinfo {year} {1963})}\BibitemShut {NoStop}%
\bibitem [{\citenamefont {Valentinis}\ \emph
  {et~al.}(2016{\natexlab{a}})\citenamefont {Valentinis}, \citenamefont
  {van~der Marel},\ and\ \citenamefont {Berthod}}]{Valentinis-2016b}%
  \BibitemOpen
  \bibfield  {author} {\bibinfo {author} {\bibfnamefont {D.}~\bibnamefont
  {Valentinis}}, \bibinfo {author} {\bibfnamefont {D.}~\bibnamefont {van~der
  Marel}}, \ and\ \bibinfo {author} {\bibfnamefont {C.}~\bibnamefont
  {Berthod}},\ }\href {\doibase 10.1103/PhysRevB.94.054516} {\bibfield
  {journal} {\bibinfo  {journal} {Phys. Rev. B}\ }\textbf {\bibinfo {volume}
  {94}},\ \bibinfo {pages} {054516} (\bibinfo {year}
  {2016}{\natexlab{a}})}\BibitemShut {NoStop}%
\bibitem [{\citenamefont {Gariglio}\ \emph {et~al.}(2016)\citenamefont
  {Gariglio}, \citenamefont {Gabay},\ and\ \citenamefont
  {Triscone}}]{Gariglio-2016}%
  \BibitemOpen
  \bibfield  {author} {\bibinfo {author} {\bibfnamefont {S.}~\bibnamefont
  {Gariglio}}, \bibinfo {author} {\bibfnamefont {M.}~\bibnamefont {Gabay}}, \
  and\ \bibinfo {author} {\bibfnamefont {J.-M.}\ \bibnamefont {Triscone}},\
  }\href {\doibase 10.1063/1.4953822} {\bibfield  {journal} {\bibinfo
  {journal} {APL Mater.}\ }\textbf {\bibinfo {volume} {4}},\ \bibinfo {pages}
  {060701} (\bibinfo {year} {2016})}\BibitemShut {NoStop}%
\bibitem [{\citenamefont {van~der Marel}\ \emph {et~al.}(2011)\citenamefont
  {van~der Marel}, \citenamefont {van Mechelen},\ and\ \citenamefont
  {Mazin}}]{vanderMarel-2011}%
  \BibitemOpen
  \bibfield  {author} {\bibinfo {author} {\bibfnamefont {D.}~\bibnamefont
  {van~der Marel}}, \bibinfo {author} {\bibfnamefont {J.~L.~M.}\ \bibnamefont
  {van Mechelen}}, \ and\ \bibinfo {author} {\bibfnamefont {I.~I.}\
  \bibnamefont {Mazin}},\ }\href {\doibase 10.1103/PhysRevB.84.205111}
  {\bibfield  {journal} {\bibinfo  {journal} {Phys. Rev. B}\ }\textbf {\bibinfo
  {volume} {84}},\ \bibinfo {pages} {205111} (\bibinfo {year}
  {2011})}\BibitemShut {NoStop}%
\bibitem [{\citenamefont {Allen}\ \emph {et~al.}(2013)\citenamefont {Allen},
  \citenamefont {Jalan}, \citenamefont {Lee}, \citenamefont {Ouellette},
  \citenamefont {Khalsa}, \citenamefont {Jaroszynski}, \citenamefont
  {Stemmer},\ and\ \citenamefont {MacDonald}}]{Allen-2013}%
  \BibitemOpen
  \bibfield  {author} {\bibinfo {author} {\bibfnamefont {S.~J.}\ \bibnamefont
  {Allen}}, \bibinfo {author} {\bibfnamefont {B.}~\bibnamefont {Jalan}},
  \bibinfo {author} {\bibfnamefont {S.}~\bibnamefont {Lee}}, \bibinfo {author}
  {\bibfnamefont {D.~G.}\ \bibnamefont {Ouellette}}, \bibinfo {author}
  {\bibfnamefont {G.}~\bibnamefont {Khalsa}}, \bibinfo {author} {\bibfnamefont
  {J.}~\bibnamefont {Jaroszynski}}, \bibinfo {author} {\bibfnamefont
  {S.}~\bibnamefont {Stemmer}}, \ and\ \bibinfo {author} {\bibfnamefont
  {A.~H.}\ \bibnamefont {MacDonald}},\ }\href 
  {\doibase 10.1103/PhysRevB.88.045114} {\bibfield  {journal} {\bibinfo  {journal} {Phys.
  Rev. B}\ }\textbf {\bibinfo {volume} {88}},\ \bibinfo {pages} {045114}
  (\bibinfo {year} {2013})}\BibitemShut {NoStop}%
\bibitem [{\citenamefont {Eagles}(1969)}]{Eagles-1969c}%
  \BibitemOpen
  \bibfield  {author} {\bibinfo {author} {\bibfnamefont {D.~M.}\ \bibnamefont
  {Eagles}},\ }\href {\doibase 10.1103/PhysRev.181.1278} {\bibfield  {journal}
  {\bibinfo  {journal} {Phys. Rev.}\ }\textbf {\bibinfo {volume} {181}},\
  \bibinfo {pages} {1278} (\bibinfo {year} {1969})}\BibitemShut {NoStop}%
\bibitem [{\citenamefont {van Mechelen}\ \emph {et~al.}(2008)\citenamefont {van
  Mechelen}, \citenamefont {van~der Marel}, \citenamefont {Grimaldi},
  \citenamefont {Kuzmenko}, \citenamefont {Armitage}, \citenamefont {Reyren},
  \citenamefont {Hagemann},\ and\ \citenamefont {Mazin}}]{vanMechelen-2008}%
  \BibitemOpen
  \bibfield  {author} {\bibinfo {author} {\bibfnamefont {J.~L.~M.}\
  \bibnamefont {van Mechelen}}, \bibinfo {author} {\bibfnamefont
  {D.}~\bibnamefont {van~der Marel}}, \bibinfo {author} {\bibfnamefont
  {C.}~\bibnamefont {Grimaldi}}, \bibinfo {author} {\bibfnamefont {A.~B.}\
  \bibnamefont {Kuzmenko}}, \bibinfo {author} {\bibfnamefont {N.~P.}\
  \bibnamefont {Armitage}}, \bibinfo {author} {\bibfnamefont {N.}~\bibnamefont
  {Reyren}}, \bibinfo {author} {\bibfnamefont {H.}~\bibnamefont {Hagemann}}, \
  and\ \bibinfo {author} {\bibfnamefont {I.~I.}\ \bibnamefont {Mazin}},\ }\href
  {\doibase 10.1103/PhysRevLett.100.226403} {\bibfield  {journal} {\bibinfo
  {journal} {Phys. Rev. Lett.}\ }\textbf {\bibinfo {volume} {100}},\ \bibinfo
  {pages} {226403} (\bibinfo {year} {2008})}\BibitemShut {NoStop}%
\bibitem [{\citenamefont {Devreese}\ \emph {et~al.}(2010)\citenamefont
  {Devreese}, \citenamefont {Klimin}, \citenamefont {van Mechelen},\ and\
  \citenamefont {van~der Marel}}]{Devreese-2010}%
  \BibitemOpen
  \bibfield  {author} {\bibinfo {author} {\bibfnamefont {J.~T.}\ \bibnamefont
  {Devreese}}, \bibinfo {author} {\bibfnamefont {S.~N.}\ \bibnamefont
  {Klimin}}, \bibinfo {author} {\bibfnamefont {J.~L.~M.}\ \bibnamefont {van
  Mechelen}}, \ and\ \bibinfo {author} {\bibfnamefont {D.}~\bibnamefont
  {van~der Marel}},\ }\href {\doibase 10.1103/PhysRevB.81.125119} {\bibfield
  {journal} {\bibinfo  {journal} {Phys. Rev. B}\ }\textbf {\bibinfo {volume}
  {81}},\ \bibinfo {pages} {125119} (\bibinfo {year} {2010})}\BibitemShut
  {NoStop}%
\bibitem [{\citenamefont {Edge}\ \emph {et~al.}(2015)\citenamefont {Edge},
  \citenamefont {Kedem}, \citenamefont {Aschauer}, \citenamefont {Spaldin},\
  and\ \citenamefont {Balatsky}}]{Edge-2015}%
  \BibitemOpen
  \bibfield  {author} {\bibinfo {author} {\bibfnamefont {J.~M.}\ \bibnamefont
  {Edge}}, \bibinfo {author} {\bibfnamefont {Y.}~\bibnamefont {Kedem}},
  \bibinfo {author} {\bibfnamefont {U.}~\bibnamefont {Aschauer}}, \bibinfo
  {author} {\bibfnamefont {N.~A.}\ \bibnamefont {Spaldin}}, \ and\ \bibinfo
  {author} {\bibfnamefont {A.~V.}\ \bibnamefont {Balatsky}},\ }\href 
  {\doibase 10.1103/PhysRevLett.115.247002} {\bibfield  {journal} {\bibinfo  {journal}
  {Phys. Rev. Lett.}\ }\textbf {\bibinfo {volume} {115}},\ \bibinfo {pages}
  {247002} (\bibinfo {year} {2015})}\BibitemShut {NoStop}%
\bibitem [{\citenamefont {Ahrens}\ \emph {et~al.}(2007)\citenamefont {Ahrens},
  \citenamefont {Merkle}, \citenamefont {Rahmati},\ and\ \citenamefont
  {Maier}}]{Ahrens-2007}%
  \BibitemOpen
  \bibfield  {author} {\bibinfo {author} {\bibfnamefont {M.}~\bibnamefont
  {Ahrens}}, \bibinfo {author} {\bibfnamefont {R.}~\bibnamefont {Merkle}},
  \bibinfo {author} {\bibfnamefont {B.}~\bibnamefont {Rahmati}}, \ and\
  \bibinfo {author} {\bibfnamefont {J.}~\bibnamefont {Maier}},\ }\href
  {\doibase 10.1016/j.physb.2007.01.008} {\bibfield  {journal} {\bibinfo
  {journal} {Physica B}\ }\textbf {\bibinfo {volume} {393}},\ \bibinfo {pages}
  {239} (\bibinfo {year} {2007})}\BibitemShut {NoStop}%
\bibitem [{\citenamefont {Fernandes}\ \emph {et~al.}(2013)\citenamefont
  {Fernandes}, \citenamefont {Haraldsen}, \citenamefont {W{\"o}lfle},\ and\
  \citenamefont {Balatsky}}]{Fernandes-2013}%
  \BibitemOpen
  \bibfield  {author} {\bibinfo {author} {\bibfnamefont {R.~M.}\ \bibnamefont
  {Fernandes}}, \bibinfo {author} {\bibfnamefont {J.~T.}\ \bibnamefont
  {Haraldsen}}, \bibinfo {author} {\bibfnamefont {P.}~\bibnamefont
  {W{\"o}lfle}}, \ and\ \bibinfo {author} {\bibfnamefont {A.~V.}\ \bibnamefont
  {Balatsky}},\ }\href {\doibase 10.1103/PhysRevB.87.014510} {\bibfield
  {journal} {\bibinfo  {journal} {Phys. Rev. B}\ }\textbf {\bibinfo {volume}
  {87}},\ \bibinfo {pages} {014510} (\bibinfo {year} {2013})}\BibitemShut
  {NoStop}%
\bibitem [{\citenamefont {Valentinis}\ \emph
  {et~al.}(2016{\natexlab{b}})\citenamefont {Valentinis}, \citenamefont
  {van~der Marel},\ and\ \citenamefont {Berthod}}]{Valentinis-2016a}%
  \BibitemOpen
  \bibfield  {author} {\bibinfo {author} {\bibfnamefont {D.}~\bibnamefont
  {Valentinis}}, \bibinfo {author} {\bibfnamefont {D.}~\bibnamefont {van~der
  Marel}}, \ and\ \bibinfo {author} {\bibfnamefont {C.}~\bibnamefont
  {Berthod}},\ }\href {\doibase 10.1103/PhysRevB.94.024511} {\bibfield
  {journal} {\bibinfo  {journal} {Phys. Rev. B}\ }\textbf {\bibinfo {volume}
  {94}},\ \bibinfo {pages} {024511} (\bibinfo {year}
  {2016}{\natexlab{b}})}\BibitemShut {NoStop}%
\bibitem [{Note1()}]{Note1}%
  \BibitemOpen
  \bibinfo {note} {In fact, the contribution of the heavy band is dominant: if
  the interaction is switched off in the light band the calculated $T_c$
  changes by less than a percent, while if the interaction in the heavy band is
  reduced by 20\%, $T_c$ does not exceed 80~mK at all densities.}\BibitemShut
  {Stop}%
\bibitem [{\citenamefont {Reyren}\ \emph {et~al.}(2009)\citenamefont {Reyren},
  \citenamefont {Gariglio}, \citenamefont {Caviglia}, \citenamefont {Jaccard},
  \citenamefont {Schneider},\ and\ \citenamefont {Triscone}}]{Reyren-2009}%
  \BibitemOpen
  \bibfield  {author} {\bibinfo {author} {\bibfnamefont {N.}~\bibnamefont
  {Reyren}}, \bibinfo {author} {\bibfnamefont {S.}~\bibnamefont {Gariglio}},
  \bibinfo {author} {\bibfnamefont {A.~D.}\ \bibnamefont {Caviglia}}, \bibinfo
  {author} {\bibfnamefont {D.}~\bibnamefont {Jaccard}}, \bibinfo {author}
  {\bibfnamefont {T.}~\bibnamefont {Schneider}}, \ and\ \bibinfo {author}
  {\bibfnamefont {J.-M.}\ \bibnamefont {Triscone}},\ }\href 
  {\doibase 10.1063/1.3100777} {\bibfield  {journal} {\bibinfo  {journal} {Appl. Phys.
  Lett.}\ }\textbf {\bibinfo {volume} {94}},\ \bibinfo {pages} {112506}
  (\bibinfo {year} {2009})}\BibitemShut {NoStop}%
\bibitem [{\citenamefont {Cancellieri}\ \emph {et~al.}(2010)\citenamefont
  {Cancellieri}, \citenamefont {Reyren}, \citenamefont {Gariglio},
  \citenamefont {Caviglia}, \citenamefont {F{\^{e}}te},\ and\ \citenamefont
  {Triscone}}]{Cancellieri-2010}%
  \BibitemOpen
  \bibfield  {author} {\bibinfo {author} {\bibfnamefont {C.}~\bibnamefont
  {Cancellieri}}, \bibinfo {author} {\bibfnamefont {R.}~\bibnamefont {Reyren}},
  \bibinfo {author} {\bibfnamefont {S.}~\bibnamefont {Gariglio}}, \bibinfo
  {author} {\bibfnamefont {A.~D.}\ \bibnamefont {Caviglia}}, \bibinfo {author}
  {\bibfnamefont {A.}~\bibnamefont {F{\^{e}}te}}, \ and\ \bibinfo {author}
  {\bibfnamefont {J.-M.}\ \bibnamefont {Triscone}},\ }\href 
  {\doibase 10.1209/0295-5075/91/17004} {\bibfield  {journal} {\bibinfo  {journal}
  {Europhys. Lett.}\ }\textbf {\bibinfo {volume} {91}},\ \bibinfo {pages}
  {17004} (\bibinfo {year} {2010})}\BibitemShut {NoStop}%
\bibitem [{\citenamefont {Gariglio}\ \emph
  {et~al.}(2015{\natexlab{b}})\citenamefont {Gariglio}, \citenamefont {Gabay},
  \citenamefont {Mannhart},\ and\ \citenamefont {Triscone}}]{Gariglio-2015a}%
  \BibitemOpen
  \bibfield  {author} {\bibinfo {author} {\bibfnamefont {S.}~\bibnamefont
  {Gariglio}}, \bibinfo {author} {\bibfnamefont {M.}~\bibnamefont {Gabay}},
  \bibinfo {author} {\bibfnamefont {J.}~\bibnamefont {Mannhart}}, \ and\
  \bibinfo {author} {\bibfnamefont {J.-M.}\ \bibnamefont {Triscone}},\ }\href
  {\doibase 10.1016/j.physc.2015.02.028} {\bibfield  {journal} {\bibinfo
  {journal} {Physica C}\ }\textbf {\bibinfo {volume} {514}},\ \bibinfo {pages}
  {189} (\bibinfo {year} {2015}{\natexlab{b}})}\BibitemShut {NoStop}%
\bibitem [{\citenamefont {Tinkham}(1996)}]{Tinkham-1996}%
  \BibitemOpen
  \bibfield  {author} {\bibinfo {author} {\bibfnamefont {M.}~\bibnamefont
  {Tinkham}},\ }\href@noop {} {\emph {\bibinfo {title} {Introduction to
  Superconductivity}}}\ (\bibinfo  {publisher} {McGraw-Hill},\ \bibinfo
  {address} {New York},\ \bibinfo {year} {1996})\BibitemShut {NoStop}%
\bibitem [{\citenamefont {Ben~Shalom}\ \emph {et~al.}(2010)\citenamefont
  {Ben~Shalom}, \citenamefont {Ron}, \citenamefont {Palevski},\ and\
  \citenamefont {Dagan}}]{BenShalom-2010b}%
  \BibitemOpen
  \bibfield  {author} {\bibinfo {author} {\bibfnamefont {M.}~\bibnamefont
  {Ben~Shalom}}, \bibinfo {author} {\bibfnamefont {A.}~\bibnamefont {Ron}},
  \bibinfo {author} {\bibfnamefont {A.}~\bibnamefont {Palevski}}, \ and\
  \bibinfo {author} {\bibfnamefont {Y.}~\bibnamefont {Dagan}},\ }\href
  {\doibase 10.1103/PhysRevLett.105.206401} {\bibfield  {journal} {\bibinfo
  {journal} {Phys. Rev. Lett.}\ }\textbf {\bibinfo {volume} {105}},\ \bibinfo
  {pages} {206401} (\bibinfo {year} {2010})}\BibitemShut {NoStop}%
\bibitem [{\citenamefont {F{\^e}te}\ \emph {et~al.}(2014)\citenamefont
  {F{\^e}te}, \citenamefont {Gariglio}, \citenamefont {Berthod}, \citenamefont
  {Li}, \citenamefont {Stornaiuolo}, \citenamefont {Gabay},\ and\ \citenamefont
  {Triscone}}]{Fete-2014}%
  \BibitemOpen
  \bibfield  {author} {\bibinfo {author} {\bibfnamefont {A.}~\bibnamefont
  {F{\^e}te}}, \bibinfo {author} {\bibfnamefont {S.}~\bibnamefont {Gariglio}},
  \bibinfo {author} {\bibfnamefont {C.}~\bibnamefont {Berthod}}, \bibinfo
  {author} {\bibfnamefont {D.}~\bibnamefont {Li}}, \bibinfo {author}
  {\bibfnamefont {D.}~\bibnamefont {Stornaiuolo}}, \bibinfo {author}
  {\bibfnamefont {M.}~\bibnamefont {Gabay}}, \ and\ \bibinfo {author}
  {\bibfnamefont {J.-M.}\ \bibnamefont {Triscone}},\ }\href 
  {\doibase 10.1088/1367-2630/16/11/112002} {\bibfield  {journal} {\bibinfo  {journal}
  {New J. Phys.}\ }\textbf {\bibinfo {volume} {16}},\ \bibinfo {pages} {112002}
  (\bibinfo {year} {2014})}\BibitemShut {NoStop}%
\bibitem [{\citenamefont {Stengel}(2011)}]{Stengel-2011}%
  \BibitemOpen
  \bibfield  {author} {\bibinfo {author} {\bibfnamefont {M.}~\bibnamefont
  {Stengel}},\ }\href {\doibase 10.1103/PhysRevLett.106.136803} {\bibfield
  {journal} {\bibinfo  {journal} {Phys. Rev. Lett.}\ }\textbf {\bibinfo
  {volume} {106}},\ \bibinfo {pages} {136803} (\bibinfo {year}
  {2011})}\BibitemShut {NoStop}%
\bibitem [{\citenamefont {Basletic}\ \emph {et~al.}(2008)\citenamefont
  {Basletic}, \citenamefont {Maurice}, \citenamefont {Carretero}, \citenamefont
  {Herranz}, \citenamefont {Copie}, \citenamefont {Bibes}, \citenamefont
  {Jacquet}, \citenamefont {Bouzehouane}, \citenamefont {Fusil},\ and\
  \citenamefont {Barthelemy}}]{Basletic-2008}%
  \BibitemOpen
  \bibfield  {author} {\bibinfo {author} {\bibfnamefont {M.}~\bibnamefont
  {Basletic}}, \bibinfo {author} {\bibfnamefont {J.~L.}\ \bibnamefont
  {Maurice}}, \bibinfo {author} {\bibfnamefont {C.}~\bibnamefont {Carretero}},
  \bibinfo {author} {\bibfnamefont {G.}~\bibnamefont {Herranz}}, \bibinfo
  {author} {\bibfnamefont {O.}~\bibnamefont {Copie}}, \bibinfo {author}
  {\bibfnamefont {M.}~\bibnamefont {Bibes}}, \bibinfo {author} {\bibfnamefont
  {E.}~\bibnamefont {Jacquet}}, \bibinfo {author} {\bibfnamefont
  {K.}~\bibnamefont {Bouzehouane}}, \bibinfo {author} {\bibfnamefont
  {S.}~\bibnamefont {Fusil}}, \ and\ \bibinfo {author} {\bibfnamefont
  {A.}~\bibnamefont {Barthelemy}},\ }\href {\doibase 10.1038/nmat2223}
  {\bibfield  {journal} {\bibinfo  {journal} {Nat. Mater.}\ }\textbf {\bibinfo
  {volume} {7}},\ \bibinfo {pages} {621} (\bibinfo {year} {2008})}\BibitemShut
  {NoStop}%
\bibitem [{\citenamefont {Cancellieri}\ \emph {et~al.}(2014)\citenamefont
  {Cancellieri}, \citenamefont {Reinle-Schmitt}, \citenamefont {Kobayashi},
  \citenamefont {Strocov}, \citenamefont {Willmott}, \citenamefont {Fontaine},
  \citenamefont {Ghosez}, \citenamefont {Filippetti}, \citenamefont {Delugas},\
  and\ \citenamefont {Fiorentini}}]{Cancellieri-2014}%
  \BibitemOpen
  \bibfield  {author} {\bibinfo {author} {\bibfnamefont {C.}~\bibnamefont
  {Cancellieri}}, \bibinfo {author} {\bibfnamefont {M.~L.}\ \bibnamefont
  {Reinle-Schmitt}}, \bibinfo {author} {\bibfnamefont {M.}~\bibnamefont
  {Kobayashi}}, \bibinfo {author} {\bibfnamefont {V.~N.}\ \bibnamefont
  {Strocov}}, \bibinfo {author} {\bibfnamefont {P.~R.}\ \bibnamefont
  {Willmott}}, \bibinfo {author} {\bibfnamefont {D.}~\bibnamefont {Fontaine}},
  \bibinfo {author} {\bibfnamefont {P.}~\bibnamefont {Ghosez}}, \bibinfo
  {author} {\bibfnamefont {A.}~\bibnamefont {Filippetti}}, \bibinfo {author}
  {\bibfnamefont {P.}~\bibnamefont {Delugas}}, \ and\ \bibinfo {author}
  {\bibfnamefont {V.}~\bibnamefont {Fiorentini}},\ }\href 
  {\doibase 10.1103/PhysRevB.89.121412} {\bibfield  {journal} {\bibinfo  {journal} {Phys.
  Rev. B}\ }\textbf {\bibinfo {volume} {89}},\ \bibinfo {pages} {121412}
  (\bibinfo {year} {2014})}\BibitemShut {NoStop}%
\bibitem [{\citenamefont {Christen}\ \emph {et~al.}(1994)\citenamefont
  {Christen}, \citenamefont {Mannhart}, \citenamefont {Williams},\ and\
  \citenamefont {Gerber}}]{Christen-1994}%
  \BibitemOpen
  \bibfield  {author} {\bibinfo {author} {\bibfnamefont {H.-M.}\ \bibnamefont
  {Christen}}, \bibinfo {author} {\bibfnamefont {J.}~\bibnamefont {Mannhart}},
  \bibinfo {author} {\bibfnamefont {E.~J.}\ \bibnamefont {Williams}}, \ and\
  \bibinfo {author} {\bibfnamefont {C.}~\bibnamefont {Gerber}},\ }\href
  {\doibase 10.1103/PhysRevB.49.12095} {\bibfield  {journal} {\bibinfo
  {journal} {Phys. Rev. B}\ }\textbf {\bibinfo {volume} {49}},\ \bibinfo
  {pages} {12095} (\bibinfo {year} {1994})}\BibitemShut {NoStop}%
\bibitem [{\citenamefont {Richter}\ \emph {et~al.}(2013)\citenamefont
  {Richter}, \citenamefont {Boschker}, \citenamefont {Dietsche}, \citenamefont
  {Fillis-Tsirakis}, \citenamefont {Jany}, \citenamefont {Loder}, \citenamefont
  {Kourkoutis}, \citenamefont {Muller}, \citenamefont {Kirtley}, \citenamefont
  {Schneider},\ and\ \citenamefont {Mannhart}}]{Richter-2013}%
  \BibitemOpen
  \bibfield  {author} {\bibinfo {author} {\bibfnamefont {C.}~\bibnamefont
  {Richter}}, \bibinfo {author} {\bibfnamefont {H.}~\bibnamefont {Boschker}},
  \bibinfo {author} {\bibfnamefont {W.}~\bibnamefont {Dietsche}}, \bibinfo
  {author} {\bibfnamefont {E.}~\bibnamefont {Fillis-Tsirakis}}, \bibinfo
  {author} {\bibfnamefont {R.}~\bibnamefont {Jany}}, \bibinfo {author}
  {\bibfnamefont {F.}~\bibnamefont {Loder}}, \bibinfo {author} {\bibfnamefont
  {L.~F.}\ \bibnamefont {Kourkoutis}}, \bibinfo {author} {\bibfnamefont
  {D.~a.}\ \bibnamefont {Muller}}, \bibinfo {author} {\bibfnamefont {J.~R.}\
  \bibnamefont {Kirtley}}, \bibinfo {author} {\bibfnamefont {C.~W.}\
  \bibnamefont {Schneider}}, \ and\ \bibinfo {author} {\bibfnamefont
  {J.}~\bibnamefont {Mannhart}},\ }\href {\doibase 10.1038/nature12494}
  {\bibfield  {journal} {\bibinfo  {journal} {Nature}\ }\textbf {\bibinfo
  {volume} {502}},\ \bibinfo {pages} {528} (\bibinfo {year}
  {2013})}\BibitemShut {NoStop}%
\end{thebibliography}

%

\end{document}